\documentclass{amsart}

\usepackage[foot]{amsaddr}
\usepackage[margin=1in]{geometry}
\usepackage{fancybox}
\usepackage{newtxtt}
\usepackage[giveninits,date=year,sortcites,maxcitenames=4]{biblatex}
\usepackage{algorithm}
\usepackage{algpseudocode}
\usepackage{bm}
\usepackage{booktabs}
\usepackage{enumitem}
\usepackage{graphicx}
\usepackage{hyperref}
\usepackage{cleveref}
\usepackage{tikz}
\usetikzlibrary{positioning}
\usetikzlibrary{calc}
\usepackage{pgfmath}
\usepackage{xspace}
\usepackage{listings}
\usepackage{pgfplots}
\usepackage{siunitx}

\pgfplotsset{compat=1.18}

\definecolor{RYB1}{RGB}{166,206,227}  \definecolor{RYB2}{RGB}{31,120,180}   \definecolor{RYB3}{RGB}{178,223,138}  \definecolor{RYB4}{RGB}{51,160,44}    \definecolor{RYB5}{RGB}{251,154,153}  \definecolor{RYB6}{RGB}{227,26,28}    \definecolor{RYB7}{RGB}{253,191,111}  \definecolor{RYB8}{RGB}{255,127,0}    \definecolor{RYB9}{RGB}{202,178,214}  \definecolor{RYB10}{RGB}{160,60,140}  \definecolor{RYB11}{RGB}{255,255,153} \definecolor{RYB12}{RGB}{177,89,40}   

\definecolor{RdPu1}{RGB}{122,1,119}   \definecolor{RdPu2}{RGB}{197,27,138}  \definecolor{RdPu3}{RGB}{247,104,161} 

\definecolor{Or}{RGB}{244,152,66}     

\definecolor{steelblue4}{HTML}{3B77AF} 

\pgfplotscreateplotcyclelist{will}{black,every mark/.append style={fill=white},mark=*\\RYB4!50!black,every mark/.append style={fill=RYB4},mark=*\\RYB6!50!black,every mark/.append style={fill=RYB6},mark=*\\RYB8!50!black,every mark/.append style={fill=RYB8},mark=*\\RYB10!50!black,every mark/.append style={fill=RYB10},mark=*\\RYB12!50!black,every mark/.append style={fill=RYB12},mark=*\\black,every mark/.append style={fill=white!50!black},mark=*\\RYB2!50!black,every mark/.append style={fill=RYB2},mark=*\\RYB4!50!black,densely dashed,every mark/.append style={fill=RYB4},mark=*\\RYB6!50!black,densely dashed,every mark/.append style={fill=RYB6},mark=*\\RYB8!50!black,densely dashed,every mark/.append style={fill=RYB8},mark=*\\RYB10!50!black,densely dashed,every mark/.append style={fill=RYB10},mark=*\\RYB12!50!black,densely dashed,every mark/.append style={fill=RYB12},mark=*\\black,densely dashed,every mark/.append style={fill=white!50!black},mark=*\\RYB2!50!black,densely dashed,every mark/.append style={fill=RYB2},mark=*\\}

\pgfplotscreateplotcyclelist{will_modulo_8}{black!50!black,every mark/.append style={fill=white},mark=*,mark size=1.2pt\\RYB4!50!black,every mark/.append style={fill=RYB4},mark=*,mark size=1.2pt\\RYB6!50!black,every mark/.append style={fill=RYB6},mark=*,mark size=1.2pt\\RYB8!50!black,every mark/.append style={fill=RYB8},mark=*,mark size=1.2pt\\RYB10!50!black,every mark/.append style={fill=RYB10},mark=*,mark size=1.2pt\\RYB12!50!black,every mark/.append style={fill=RYB12},mark=*,mark size=1.2pt\\black!50!black,every mark/.append style={fill=white!50!black},mark=*,mark size=1.2pt\\RYB2!50!black,every mark/.append style={fill=RYB2},mark=*,mark size=1.2pt\\black!50!black,every mark/.append style={fill=white},mark=*,mark size=0.5pt\\RYB4!70!black,every mark/.append style={fill=RYB4},mark=*,mark size=0.5pt\\RYB6!50!black!30,every mark/.append style={fill=RYB6},mark=*,mark size=0.5pt\\RYB8!50!black,every mark/.append style={fill=RYB8},mark=*,mark size=0.5pt\\RYB10!50!black,every mark/.append style={fill=RYB10},mark=*,mark size=0.5pt\\RYB12!50!black,every mark/.append style={fill=RYB12},mark=*,mark size=0.5pt\\black!50!black,every mark/.append style={fill=white!50!black},mark=*,mark size=0.5pt\\RYB2!50!black,every mark/.append style={fill=RYB2},mark=*,mark size=0.5pt\\}

\pgfplotscreateplotcyclelist{will_modulo_6}{black!50!black,every mark/.append style={fill=white},mark=*,mark size=1.2pt\\RYB4!50!black,every mark/.append style={fill=RYB4},mark=*,mark size=1.2pt\\RYB6!50!black,every mark/.append style={fill=RYB6},mark=*,mark size=1.2pt\\RYB8!50!black,every mark/.append style={fill=RYB8},mark=*,mark size=1.2pt\\RYB10!50!black,every mark/.append style={fill=RYB10},mark=*,mark size=1.2pt\\RYB12!50!black,every mark/.append style={fill=RYB12},mark=*,mark size=1.2pt\\black!50!black,every mark/.append style={fill=white},mark=*,mark size=0.5pt\\RYB4!70!black,every mark/.append style={fill=RYB4},mark=*,mark size=0.5pt\\RYB6!50!black!30,every mark/.append style={fill=RYB6},mark=*,mark size=0.5pt\\RYB8!50!black,every mark/.append style={fill=RYB8},mark=*,mark size=0.5pt\\RYB10!50!black,every mark/.append style={fill=RYB10},mark=*,mark size=0.5pt\\RYB12!50!black,every mark/.append style={fill=RYB12},mark=*,mark size=0.5pt\\}

\lstdefinestyle{cpp}{
    commentstyle=\color{black!50!white},
    basicstyle=\ttfamily\scriptsize,
    breakatwhitespace=true,
    breaklines=true,
    texcl=true,
}
\definecolor{code}{rgb}{0.7, 0, 0.4}
\newcommand{\code}[1]{\texttt{\small\color{code} #1}}
 
\begin{document}

\title{High-performance finite elements with MFEM}

\author{Julian Andrej$^1$}
\author{Nabil Atallah$^1$}
\author{Jan-Phillip B{\"a}cker$^2$}
\author{John Camier$^1$}
\author{Dylan Copeland$^1$}
\author{Veselin Dobrev$^1$}
\author{Yohann Dudouit$^1$}
\author{Tobias Duswald$^3$}
\author{Brendan Keith$^4$}
\author{Dohyun Kim$^4$}
\author{Tzanio Kolev$^1$}
\author{Boyan Lazarov$^5$}
\author{Ketan Mittal$^1$}
\author{Will Pazner$^{*6}$}
\author{Socratis Petrides$^1$}
\author{Syun'ichi Shiraiwa$^7$}
\author{Mark Stowell$^1$}
\author{Vladimir Tomov$^1$}

\address{$^1$Center for Applied Scientific Computing, Lawrence Livermore National Laboratory, Livermore, CA}
\address{$^2$Institute of Applied Mathematics (LS III), TU Dortmund University, Dortmund, Germany}
\address{$^3$CERN and Technical University of Munich, Munich, Germany}
\address{$^4$Division of Applied Mathematics, Brown University, Providence, RI}
\address{$^5$Computational Engineering Division, Lawrence Livermore National Laboratory, Livermore, CA}
\address{$^6$Fariborz Maseeh Department of Mathematics and Statistics, Portland State University, Portland, OR }
\address{$^7$Princeton Plasma Physics Laboratory, Princeton, NJ}

\address{$^*$Corresponding author: Will Pazner, \textup{pazner@pdx.edu}}

\begin{abstract}
   The MFEM (Modular Finite Element Methods) library is a high-performance C++ library for finite element discretizations.
   MFEM supports numerous types of finite element methods and is the discretization engine powering many computational physics and engineering applications across a number of domains.
   This paper describes some of the recent research and development in MFEM, focusing on performance portability across leadership-class supercomputing facilities, including exascale supercomputers, as well as new capabilities and functionality, enabling a wider range of applications.
   Much of this work was undertaken as part of the Department of Energy's Exascale Computing Project (ECP) in collaboration with the Center for Efficient Exascale Discretizations (CEED).
\end{abstract}

\maketitle

\section{Introduction}
The MFEM (Modular Finite Element Methods) library is a high-performance, scalable, open-source, C++ library for finite element discretizations \cite{MFEM}.
MFEM has undergone continual, rapid development since its initial release in July 2010.
Its feature set has grown significantly, supporting a large number of applications, discretizations, and target architectures.
This paper aims to summarize the recent developments in the library, as well as the related research into the underlying mathematical and numerical methods.
It can be seen as a follow-up to \cite{Anderson2020} and we recommend reviewing that paper for relevant background material.

Given the importance of graphics processing units (GPUs) to high-performance computing, MFEM's GPU capabilities have been considerably expanded and improved.
For example, MFEM now provides comprehensive support for GPU-accelerated high-order mesh optimization through the target-matrix optimization paradigm.
Additionally, MFEM provides end-to-end GPU-accelerated solvers for problems posed in all spaces of the finite element de Rham complex, making use of newly developed low-order-refined preconditioning techniques.
These solvers have recently been extended to saddle-point problems, include grad-div problems in $H(\operatorname{div})$ and mixed finite element problems.
Novel kernel fusion techniques implemented in MFEM can be used to substantially improve the strong scalability of high-order finite element solvers, resulting in peak performance on problems of sizes 5--10$\times$ smaller than using traditional approaches.
GPU-accelerated partial assembly kernels have been expanded to cover a very broad range of operators and discretizations.
For example, NURBS-based discretizations (such as isogeometric analysis) now also support partial assembly.

In addition to the performance-focused features and developments described above, MFEM's feature-set has been significantly broadened.
Support for specialized discretizations such as the discontinuous Petrov--Galerkin (DPG) method have been added.
Automatic differentiation now enables the high-performance automated assembly of Jacobian matrices for applications such as nonlinear elasticity.
Multiphysics applications are supported through submesh capabilities, providing users with straightforward mechanisms to use different physics in different parts of the domain.
Discretization techniques for stochastic and fractional partial differential equations are enabled by the library, and illustrated in MFEM's included examples and miniapps.
In addition to supporting traditional body-fitted meshes and finite element methods, MFEM now also supports level-set based methods;
specific examples include distance solvers, shifted boundary methods, and cut integration rules.

Each of these topics is discussed in greater detail in the following sections.
We begin with a summary of MFEM's GPU and HPC capabilities, including scalable matrix-free solvers.
Then, we describe advances in MFEM's discretization support, such as DPG support, proximal Galerkin methods, immersed discretizations, NURBS/IGA, and automatic differentiation.
Subsequently, we give an overview of advances in MFEM's meshing capabilities, with a focus on high-order mesh optimization algorithms.
Finally, we illustrate the utility of these features and developments by describing a number of applications that make use of the newly introduced capabilities.
 
\section{GPU Acceleration and High-Performance Computing}

One of MFEM's distinguishing features is its support for GPU acceleration throughout the library.
GPU support was introduced in MFEM version 4.0, released in May 2019.
Since then, an increasing number of features and discretizations support GPU acceleration.
For example, MFEM's powerful mesh optimization features based on the target-matrix optimization paradigm now run on GPUs, enabling over $30\times$ speedup compared to CPU evaluation.
These advanced meshing capabilities are discussed in greater detail in the meshing section of the paper.
Much of MFEM's GPU capabilities are focused on high-order discretizations.
The MFEM team continues to perform research and development in generating optimized kernels and in reaching peak performance faster in high-order finite element simulations.
A major area of research and development is that of matrix-free solvers, which are tailored for high-order finite element problems.
MFEM's most recent release includes support for full end-to-end GPU acceleration of matrix-free solvers for all spaces in the finite element de Rham complex.
These solvers are discussed in the next section.

\subsection{Matrix-free solvers for high-order finite elements}
High-order finite element problems are typically solved using an iterative method, such as a Krylov subspace method, together with an effective preconditioner.
Applying the action of high-order finite element operators in partially assembled form is very well-suited for GPU acceleration;
the arithmetic intensity of the algorithm increases with increasing polynomial degree \cite{Abdelfattah2021,Kolev2021a}.
However, the construction and application of performant preconditioners is more challenging.
It is typically prohibitively expensive (both in terms of memory usage and computational complexity) to assemble and store the system matrix associated with the discretization, see Figure \ref{fig_feod} and \cite{Anderson2020}.
As a result, traditional matrix-based methods such as algebraic multigrid cannot be directly used in this context.
Moreover, sparse matrix computations have low arithmetic intensity and are memory bandwidth bound;
these operations often do not fully utilize the computational resources afforded by GPUs.

To address these issues, MFEM supports a range of \textit{matrix-free} preconditioning methods for a variety of finite element operators and discretizations.
Such preconditioners can be constructed and applied without access to the assembled matrix, and their memory usage is asymptotically optimal (linear scaling with the problem size).
The two main approaches supported by MFEM are $p$-multigrid and low-order-refined preconditioning.

Both of these approaches aim to reduce the high-order problem to a related lower-order problem, which can then be assembled, and treated with, for example, algebraic multigrid.
In $p$-multigrid, a hierarchy of finite element spaces with different polynomial degrees is constructed on the same mesh.
These spaces are nested: there exists a natural injection from the lower-degree spaces into the higher-degree spaces.
This gives rise to restriction and prolongation operators that transfer solutions and residuals between the spaces;
the action of these operators can be computed efficiently on the GPU using sum-factorization techniques.
At each level, a smoother is required;
since the assembled matrix is not available, it is not feasible to use matrix-based relaxation methods such as Gauss--Seidel.
Instead, smoothers based only on the diagonal of the matrix are used;
sum-factorization techniques provide algorithms to efficiently compute the diagonal of finite element matrices without assembling the matrix.
Chebyshev acceleration can be used to improve the smoother.

Low-order-refined (LOR) preconditioning, also known in the literature as SEM--FEM preconditioning (see e.g.~\cite{Fischer1997}), is based on the idea of constructing a \textit{spectrally equivalent} lowest-order discretization on a refined version of the mesh.
The high-order discretization on the original mesh and the low-order discretization on the refined mesh have the same number of degrees of freedom.
Under some conditions on the construction of the mesh, these two discretization are spectrally equivalent, independent of mesh size and polynomial degree \cite{Canuto1994}.
Consequently, a preconditioner constructed using the low-order matrix can be used to effectively precondition the high-order operator.
While this technique is by now considered classical (having originally been proposed by Orszag in 1980, \cite{Orszag1980}), recent developments have extended the applicability of LOR approaches to additional discretizations, including all spaces in the finite element de Rham complex \cite{Pazner2023}, as well as discontinuous Galerkin discretizations with $hp$-refinement \cite{Pazner2020a,Pazner2021b}.

MFEM provides extensive, easy-to-use support for GPU-accelerated low-order-refined discretizations and solvers.
Spectrally equivalent low-order discretizations can be created in a single line of code.
Any of MFEM's wide array of supported preconditioners can be used for the LOR preconditioner;
for example, \textit{hypre}'s high-performance AMG, AMS, and ADS preconditioners, which are readily available in MFEM, provide effective solvers for diffusion problems posed in $H^1$, $H(\operatorname{curl})$, and $H(\operatorname{div})$.

The entire preconditioning pipeline is GPU-accelerated.
The lowest-order matrix is assembled using a batched macro-element algorithm that takes advantage of the semi-structured nature of the refined mesh.
High-performance, scalable algebraic multigrid methods can then be created directly on-device and combined with MFEM's highly performant partial assembly kernels for operator evaluation \cite{Pazner2023a}.
These solvers are highly performant and scalable up to thousands of GPUs.

Spectrally equivalent LOR discretizations in $H(\operatorname{curl})$ and $H(\operatorname{div})$ use a particular choice of basis constructed using interpolation and histopolation operators;
in this basis, the discrete curl and divergence differential operators defined on the high-order and low-order-refined spaces are exactly equal.
This remarkable property is the key fact behind the spectral equivalence of the high-order and low-order operators.
It can also be used in other contexts to create efficient preconditioners.
High-performant preconditioners for saddle-point problems resulting from mixed finite element discretizations (e.g.\ Darcy problems) can be constructed using this technique \cite{Pazner2024}.

The low-order-refined approach can also be used to couple high-order and low-order discretizations in a multiphysics or mixed discretization framework.
MFEM provides conservative and accurate transfer operators that can be used to communicate solutions and residuals between spaces of different orders and refinements \cite{Kolev2022}.
 
\subsection{State-of-the-art performance on exascale platforms and AMD GPUs}
In order to achieve high-performance on a range of GPU-based platforms, MFEM supports a spectrum of \textit{operator assembly levels}, ranging from \textit{full assembly} of the system matrix, to completely \textit{matrix-free} (zero storage) operators.
The supported assembly levels are summarized below.
Each assembly level brings its own computational efficiency and memory usage properties
(see also Figure \ref{fig_feod} and \cite{Anderson2020}).

\begin{itemize}[leftmargin=0pt,label={}]
    \item \textit{Fully Matrix-Free.} All necessary data needed for operator evaluation is computed on the fly, minimizing storage and memory requirements, at the cost of potentially recomputing data.
    \item \textit{Partial Assembly.} Only the essential data at quadrature points is precomputed and stored, decreasing computation costs compared with fully matrix-free, while reducing memory footprint compared with matrix assembly. Partial assembly is especially effective for high-order finite elements.
    \item \textit{Element Assembly.} Local contributions from each element are computed and stored in an element-local dense matrix. While this approach is more computationally and memory-intensive than partial assembly, it also gives access to an algebraic representation of the operator.
    \item \textit{Sparse Matrix Assembly.} The traditional assembly method computes and stores the global sparse matrix representing the operator. While this can be memory-intensive for large-scale problems (especially with higher order finite elements), it is compatible with a wide range of existing solvers and preconditioners.
\end{itemize}

All of the above-described assembly levels in MFEM are designed to be compatible with GPU-based execution, leveraging the computational power of modern GPUs to accelerate simulations.
This compatibility with device execution on GPUs allows MFEM to handle complex, large-scale problems more efficiently, making it a versatile tool for scientific computing and engineering simulations.

\begin{figure}
    \centering
    \includegraphics[width=\linewidth]{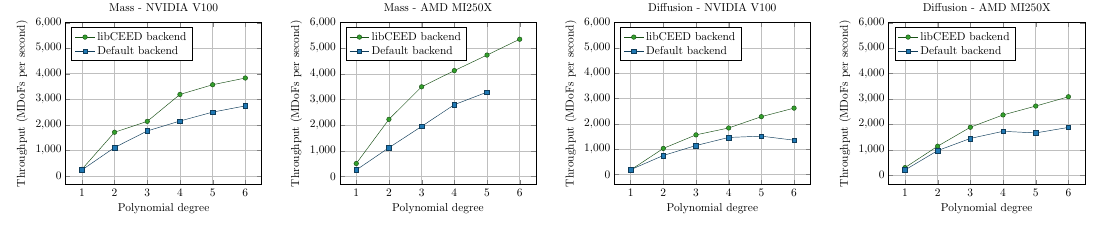}
    \caption{Comparison of the performance on the mass and diffusion benchmark problems (CEED BP1 and BP3) on NVIDIA V100 and AMD MI250X using the libCEED and default MFEM backends.}
    \label{nvidia-vs-amd}
\end{figure}

MFEM has expanded its capabilities to fully support exascale computing platforms, including optimized support for multiple GPU architectures.
The integration with AMD GPUs is particularly noteworthy, as it aligns with the increasing adoption of AMD hardware in exascale and supercomputing platforms.
Several bake-off problems (BPs) have been released by the CEED project and are used as important benchmarks: there are the mass and diffusion benchmarks: BP1 and BP3 \cite{ceed2020scalability} as well as the more recent preconditioned Poisson solvers BPS3 \cite{ceed-ms37}.
The MFEM matrix-free kernels have shown good performance portability between NVIDIA and AMD architectures, as illustrated in \Cref{nvidia-vs-amd}.
 
\subsection{Kernel fusion}
GPU kernel fusion is an important technique for maximizing performance and scalability of GPU-accelerated finite element options.
One of the challenges with it is providing a user-friendly programming interface to generate fused computational kernels.
In this section we present some of our research in this area performed in MFEM as part of the CEED project.

Coding for kernel fusion often requires a paradigm shift, moving away from explicit GPU kernel launches.
One general way to address this challenge is to gain in abstraction, using, for example, a more descriptive mathematical language.
For finite element methods, the FEniCS project has formalized the Unified Form Language (UFL, see \cite{Logg2012}).
UFL is a domain specific language used to declare finite element variational forms, thus providing a high-level abstraction for specifying the mesh, the finite element spaces, the boundary conditions, as well as the linear and bilinear forms.
Working at the compiler frontend level allows one to build a modular toolchain that transforms UFL to C++ and raw CUDA.
This enables the construction of a source-to-source transformation which uses the graph of all the kernels that are needed at runtime, together with the memory locations that are to be read, written or copied, allowing static analysis and optimizations.

The implementation of the CEED benchmark problems using kernel fusion requires several building blocks.
It is necessary to launch kernels with different topologies (e.g.~1D kernels for vector operations, and 2D and 3D thread blocks kernels for the main partially assembled operators).
Specific warp levels instructions are also needed for the dot products.
A mechanism to synchronize these different algorithmic parts is also necessary.
One natural way  to achieve these goals on Nvidia GPUs is to use the \textit{Cooperative Groups} programming model.
We use group partitioning for organization and used the group collectives for synchronizations.

\begin{figure}
   \centering
   \includegraphics[width=0.25\linewidth]{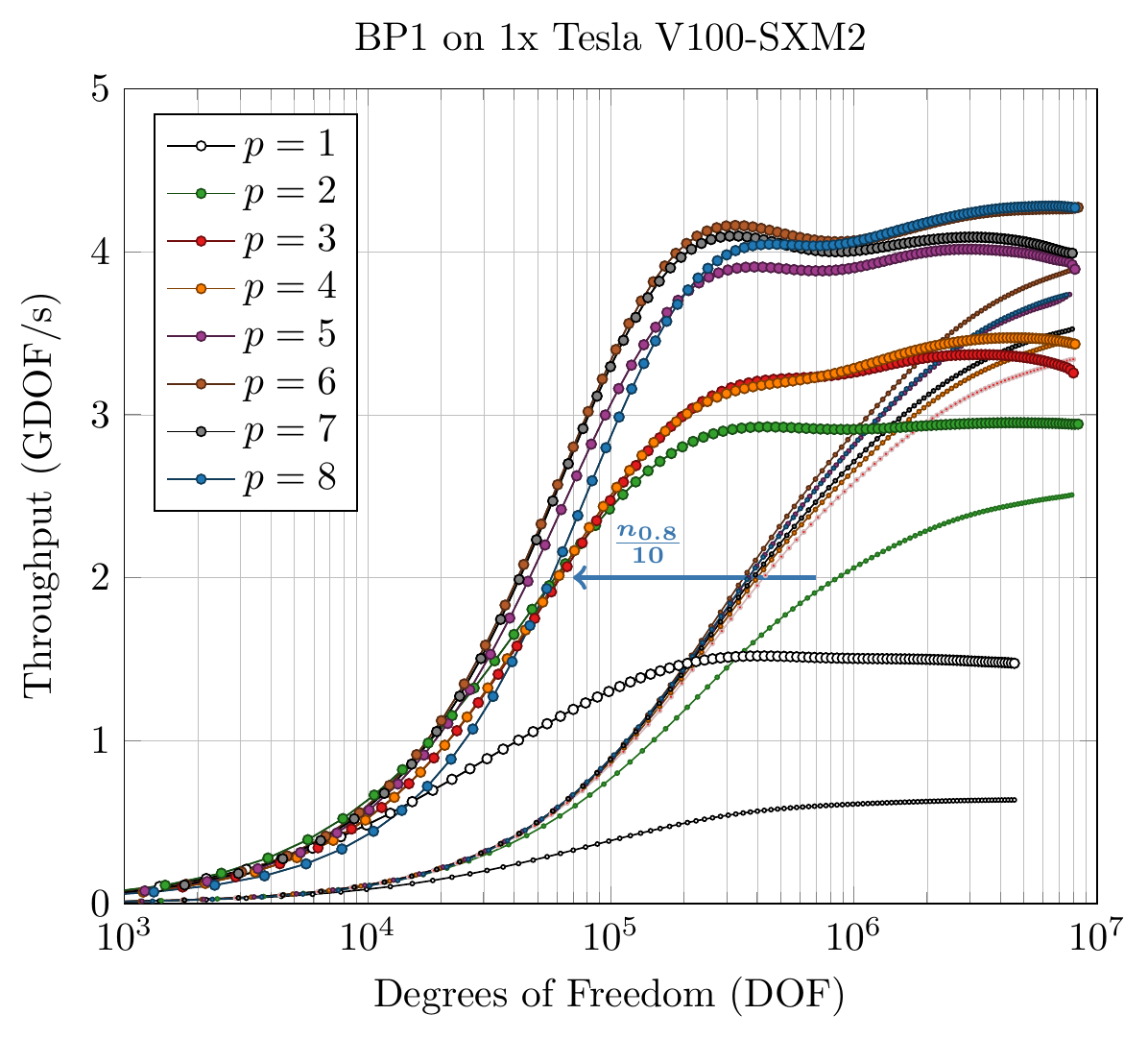}\includegraphics[width=0.25\linewidth]{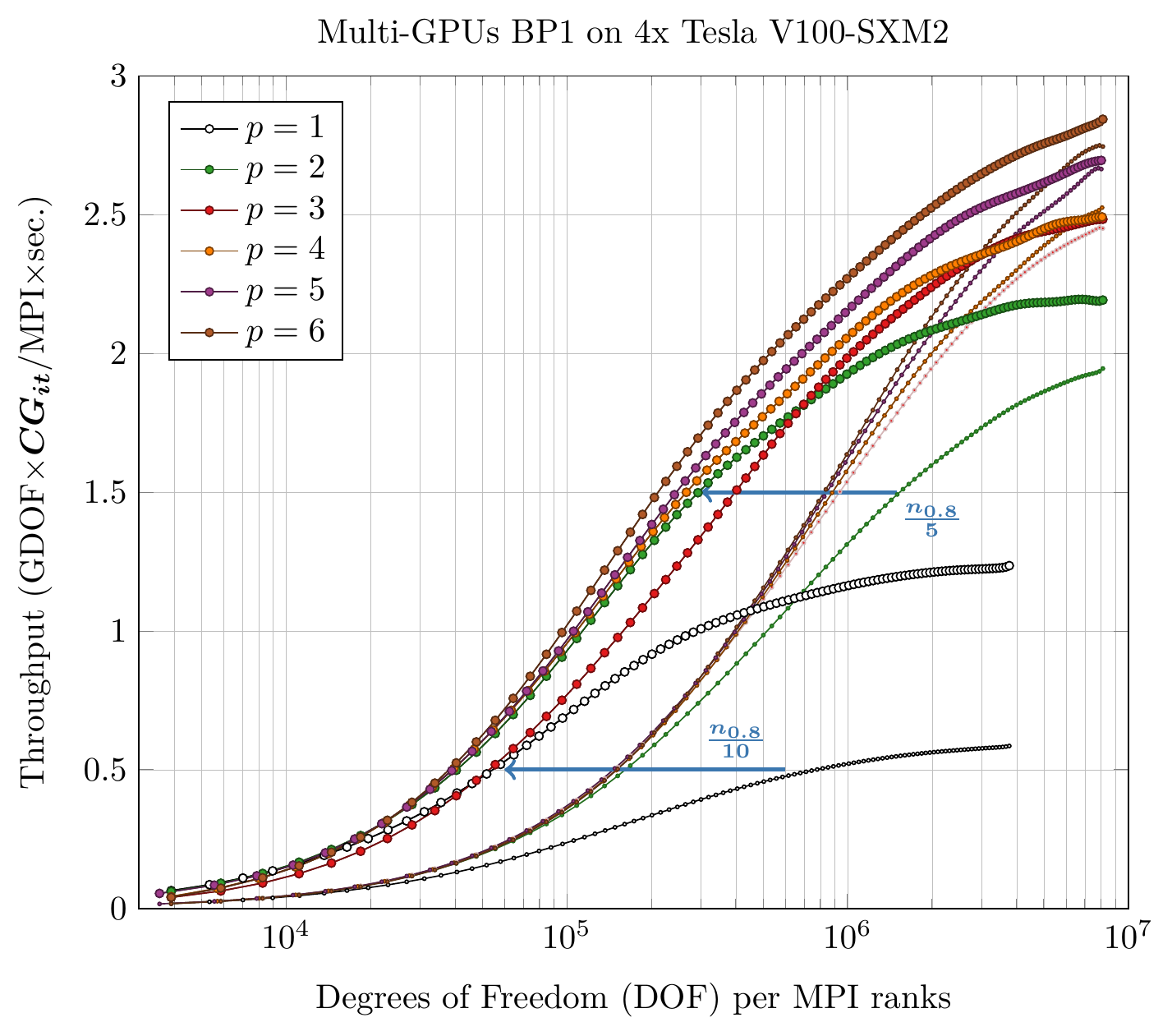}
   \caption{Throughput for fused kernel mass matrix benchmark}
   \label{fig_mfem_fusion_bp1}
\end{figure}

The results presented in the left panel of \Cref{fig_mfem_fusion_bp1} demonstrate single-GPU performance on the CEED mass matrix benchmark BP1.
One main question is whether or not such gain in performance still holds for larger problems or solvers.
Moving from a serial operator to the parallel version requires a triple product of the form $P^{T}AP$:
the finite element operator $A$ is prefixed and postfixed by multi-GPU communication operators $P$ and $P^T$.
The prefix communication operator prepares the buffers, sends the required data asynchronously to the neighbors, performs works on local vectors, and finally unpacks the received data to finalize the operation.
The transposed operator performs the same operations but in reverse order.
These methods have been implemented on NVIDIA GPUs with the \textit{NVSHMEM} library, which proposes a parallel API based on \textit{OpenSHMEM}, providing a global address space for multiple GPUs, allowing fine-grained kernel-initiated communication operations.
This approach allows us to move from standard GPU-to-CPU data transfers with the MPI interconnect model to a direct GPU-to-GPU model.
The right panel of \Cref{fig_mfem_fusion_bp1} shows the latest results obtained with a fully fused kernel with MPI communications for the CEED mass matrix benchmark.
Note that the number of degrees of freedom needed to reach 80\% of peak performance has been reduced by a factor of five for the higher orders, thereby greatly increasing the strong scalability of these operators.
To the best of our knowledge, this is the first time such an improvement has been observed with MPI communications overhead.

\subsection{Mixed meshes, nonconforming meshes, and \texorpdfstring{$p$}{p}-adaptivity}
\subsubsection{Matrix-free discontinuous Galerkin methods on non-conforming meshes.}
\begin{figure}
    \centering
    \includegraphics[width=0.5\linewidth]{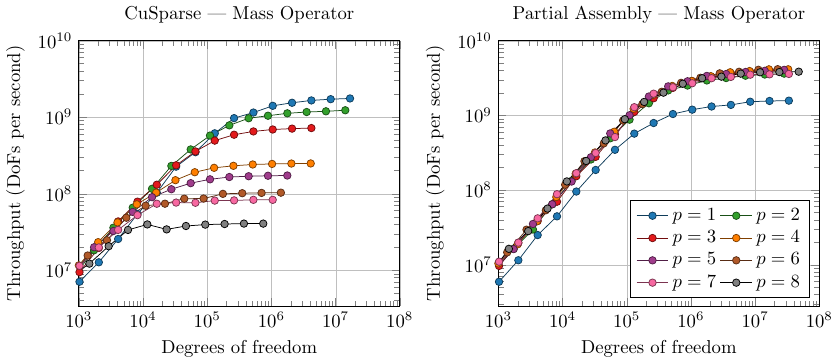}\includegraphics[width=0.5\linewidth]{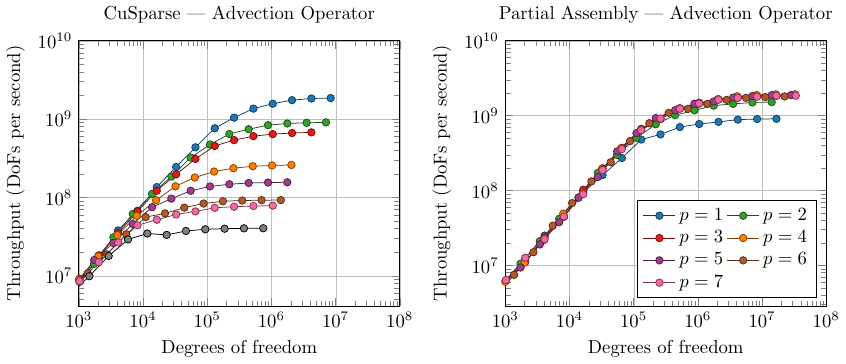}
    \caption{Performance comparison of matrix-free discontinuous Galerkin mass operators and advection operators (NVIDIA V100).}
    \label{fig:dgpa-perf}
\end{figure}

MFEM's GPU support has been enhanced with the implementation of matrix-free discontinuous Galerkin method on adaptively refined non-conforming meshes.
The non-conforming support builds on the previous implementation of conforming matrix-free discontinuous Galerkin method.
A key aspect of this implementation is the effective handling of non-conforming faces, achieved through a transformation of degrees of freedom, which allows the use of standard conforming face kernels on non-conforming meshes.
This non-conforming transformation happens at the E-vector level only on non-conforming faces, minimizing its computational cost.
For non-conforming faces, degrees of freedom are interpolated from the coarse face to the fine face such that the face appears as a conforming face from a computational perspective.
Performance results are shown in \Cref{fig:dgpa-perf}.

\subsubsection{Mixed meshes and \texorpdfstring{$p$}{p}-adaptivity.}
\begin{figure}
    \centering
    \begin{tikzpicture}[scale=0.55]
\begin{axis}[
grid=both,
domain=1:8,
  width=3in,
xlabel={Polynomial degree},
  xtick={1,2,...,6},
  ytick={0,1000,...,6000},
  ylabel={Throughput (MDoFs per second)},
  cycle list name=will,
  legend cell align=left,
  legend pos=north west,
  mark size=2pt,
legend entries={/gpu/hip/shared,/gpu/hip/magma/det,/gpu/hip/gen},
  title={MI250X - Hexahedral mesh},
  ymax=6000,
  ymin=0
]
\addplot
table {1  221.778
2 666.691
3 1053.81
4 1381.2
5 1564.95
6 1894.62
};
\addplot
table {1 208.969
2 653.153
3 970.383
4 1079.59
5 1336.69
6 1502.99
};
\addplot
table {1  508.656
2  2230.4
3 3498.66
4 4128.41
5  4735.3
6 5352.78
};
\end{axis}
\end{tikzpicture}
\qquad
\begin{tikzpicture}[scale=0.55]
\begin{axis}[
grid=both,
domain=1:8,
  width=3in,
xlabel={Polynomial degree},
  xtick={1,2,...,6},
  ytick={0,1000,...,6000},
  ylabel={Throughput (MDoFs per second)},
  cycle list name=will,
  legend cell align=left,
  legend pos=north west,
  mark size=2pt,
legend entries={/gpu/hip/shared,/gpu/hip/magma/det},
  title={MI250X - Tetrahedral mesh},
  ymax=6000,
  ymin=0
]
\addplot
table {1 53.2622
2 242.456
3 360.306
4 382.713
5 315.31
6 107.07
};
\addplot
table {1 220.905
2 806.413
3 1134.13
4 1287.45
5 1206.13
6 788.66
};
\end{axis}
\end{tikzpicture}
\qquad
\begin{tikzpicture}[scale=0.55]
\begin{axis}[
grid=both,
domain=1:8,
  width=3in,
xlabel={Polynomial degree},
  xtick={1,2,...,6},
  ytick={0,1000,...,6000},
  ylabel={Throughput (MDoFs per second)},
  cycle list name=will,
  legend cell align=left,
  legend pos=north west,
  mark size=2pt,
legend entries={/gpu/hip/shared,/gpu/hip/magma/det},
  title={MI250X - Mixed mesh},
  ymax=6000,
  ymin=0
]
\addplot
table {1 67.9605
2 179.247
3 238.628
4 168.745
5 115.693
6 106.335
};
\addplot
table {1 147.745
2 514.851
3 749.729
4 813.257
5 925.857
6 939.181
};
\end{axis}
\end{tikzpicture}
     \caption{Performance of a matrix-free mass operator (BP1 benchmark problem) for different type of meshes using an AMD MI250X GPU.}
    \label{mixed-mesh}
\end{figure}
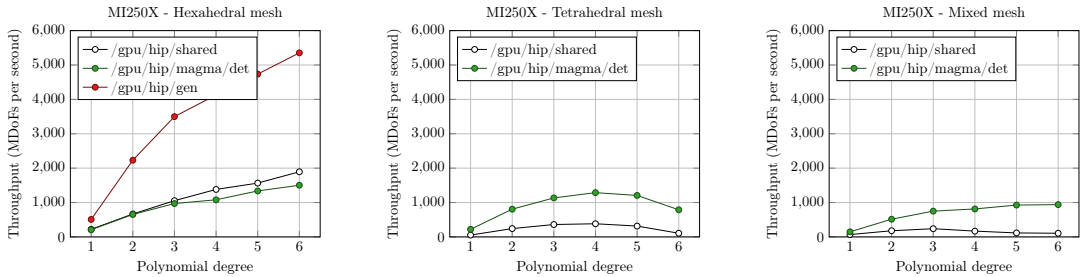

The integration of the libCEED library (see~\cite{Brown2021}) in MFEM adds unique features to matrix-free operators, including support for simplices, SYCL-based hardware, and fully matrix-free operators (zero storage operators).
The most notable recent addition is the support for mixed meshes with element of different geometric type (including simplices, pyramids, and wedges), and $p$-adaptivity.
MFEM users can run simulations using matrix-free operators on mixed meshes and $p$-adaptivity by simply using the libCEED backend.
The performance on such meshes is illustrated on Figure~\ref{mixed-mesh}.
We observe similar performance on mixed meshes and on tetrahedral meshes, however pure hexahedral meshes result in significantly higher performance due to the use of sum-factorization algorithms.
 
\section{Discretizations}

The MFEM library supports a wide range of finite element methods and discretizations.
In addition to traditional conforming finite element methods, MFEM provides ever-increasing support for alternative or non-standard finite element methods some of which are described below.
These methods differ in important ways from standard methods, and can provide advantages such as discrete stability, pointwise bounds preservation, or ability to handle implicitly defined geometries.

\subsection{Discontinuous Petrov-Galerkin (DPG) methods}

MFEM now supports a wide range of DPG formulations spanning the whole de Rham complex \cite{carstensen2016breaking}.
DPG is a non-standard minimum residual method that provides high accuracy, unconditional discrete stability and positive definite linear systems \cite{demkowicz2010class}.
It is well-suited for challenging problems that are prone to loss of discrete stability (e.g.~high-frequency time-harmonic acoustics and electromagnetic equations) and for problems that require adaptive mesh refinement (AMR) (e.g.~problems with singular solutions; see \cite{demkowicz2012class,petrides2017adaptive,petrides2021adaptive}).

\begin{figure}
   \centering
   \includegraphics[width=0.2\linewidth]{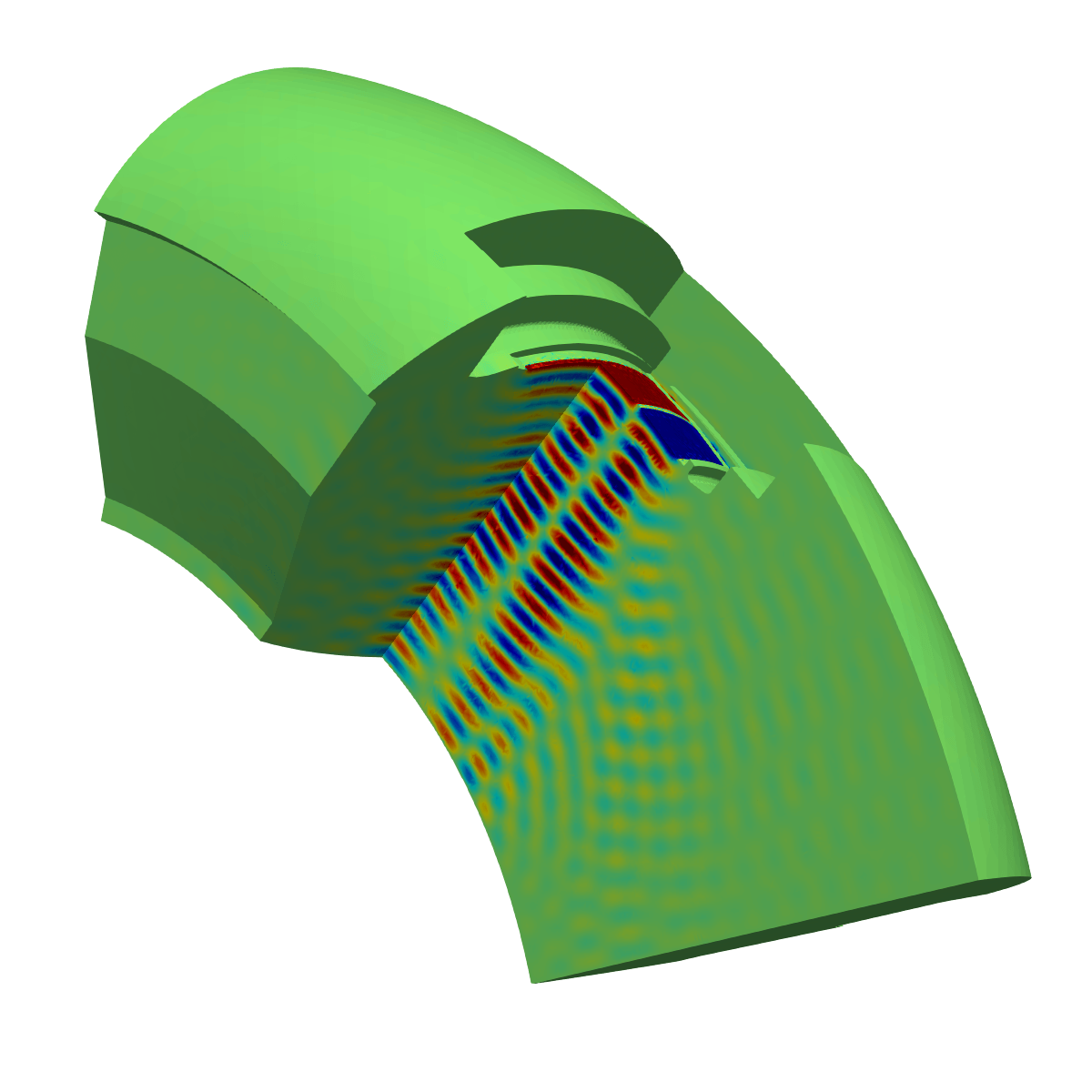}\includegraphics[width=0.2\linewidth]{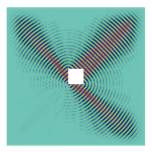}\includegraphics[width=0.2\linewidth]{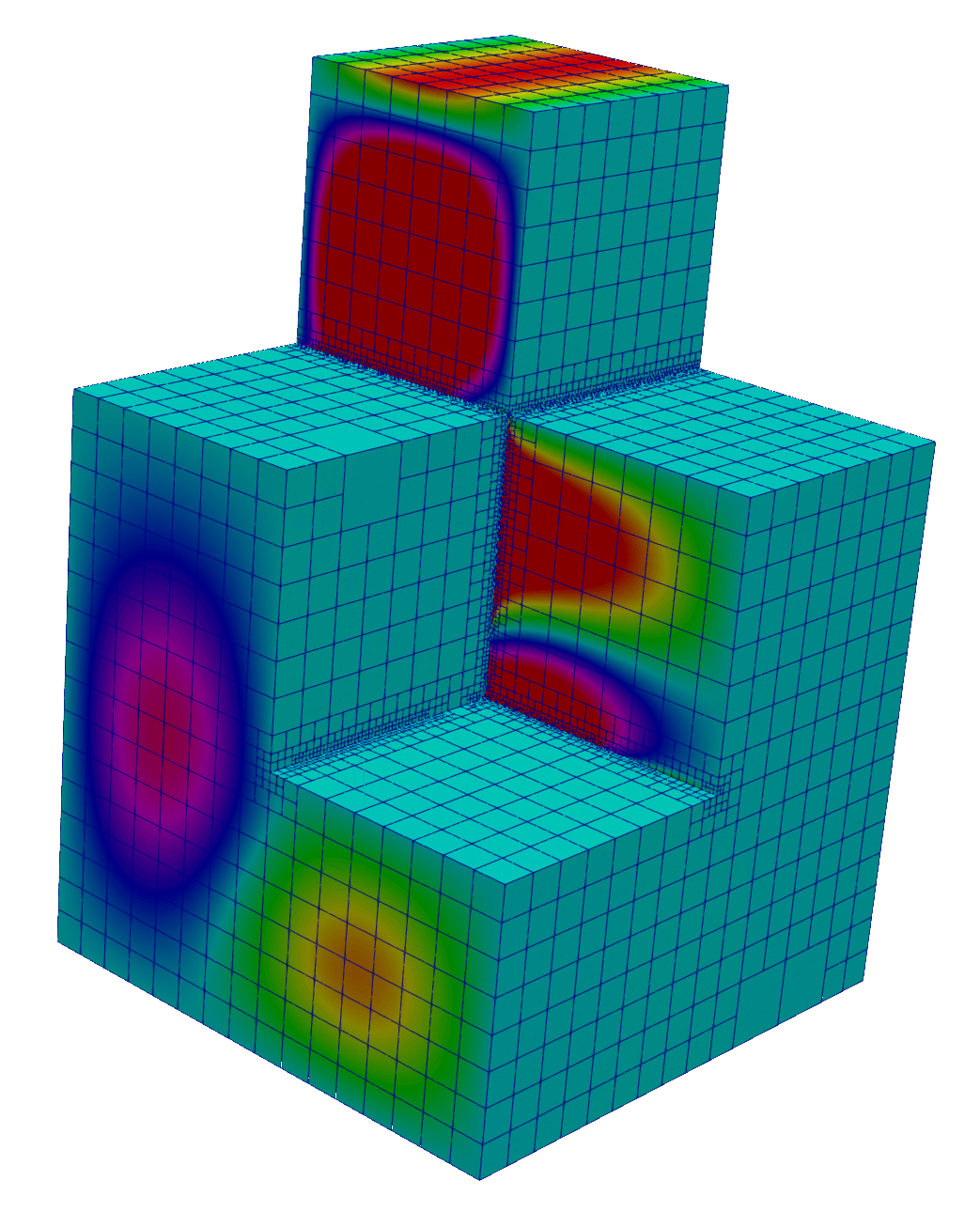}

   \caption{Ultraweak DPG formulations for time-harmonic Maxwell equations for Tokamak simulations(left),
   2D scattering of an acoustics beam (center), and for the time-harmonic Maxwell equations with AMR for the simulation of the microwave oven problem (right).}
   \label{dpg_figs}
\end{figure}

This new feature introduces user-friendly classes that allow users to define multiple trial and test spaces as well as the desired bilinear and linear integrators in a similar fashion to the standard Galerkin formulation.
The DPG linear system assembly, which also involves the on-the-fly computation of optimal test functions (see \cite{demkowicz2011class,zitelli2011class}), is then performed automatically under the hood.
The system is returned to the user in blocked form.
Complex-valued systems, static condensation and AMR driven by the DPG built-in error indicator are also supported \cite{demkowicz2012class}.
This new development is demonstrated with several examples for diffusion, convection-diffusion, time-harmonic acoustic and electromagnetic wave equations.
 
\subsection{Proximal Galerkin}

The latent variable proximal Galerkin (PG) finite element method \cite{keith2023proximal} is a high-order, nonlinear finite element method that preserves the intrinsic geometric structure of pointwise bound constraints in function spaces.
MFEM now provides implementations of PG through two new examples:
Example 36, which uses PG to solve the classical obstacle problem, is a template for other unilateral-constrained variational inequalities and free boundary problems in Sobolev spaces; and Example 37, which uses a novel information-geometric approach to topology optimization, and provides a template for using PG to solve bilateral-constrained variational inequalities and pointwise bound-constrained optimization problems in Lebesgue spaces.

\subsection{Immersed discretizations}
Capabilities for finite element calculations over immersed meshes have recently been added to the library.
These methods assume the feature of interest (domain boundary or internal interface) is given implicitly by the zero level set of a discrete function.
The new features share a common objective: adhering to finite element operations while avoiding purely geometric computations. This principle enables better generality across dimensions, element orders, and types of finite element discretization.

\subsubsection{Integration over cut elements.}

\begin{figure}
   \begin{center}
   \includegraphics[height=0.2\linewidth]{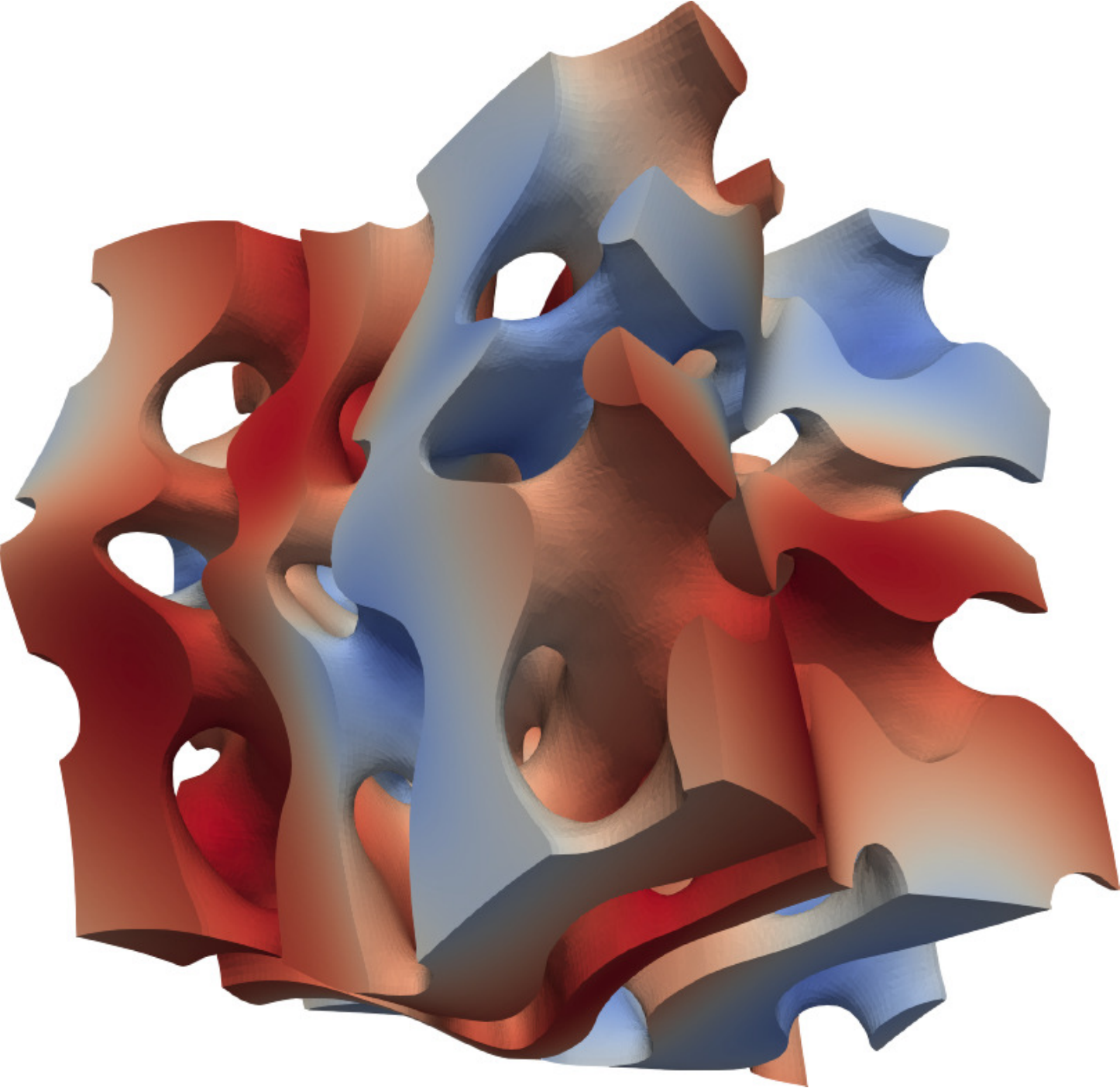}
   \includegraphics[height=0.2\linewidth]{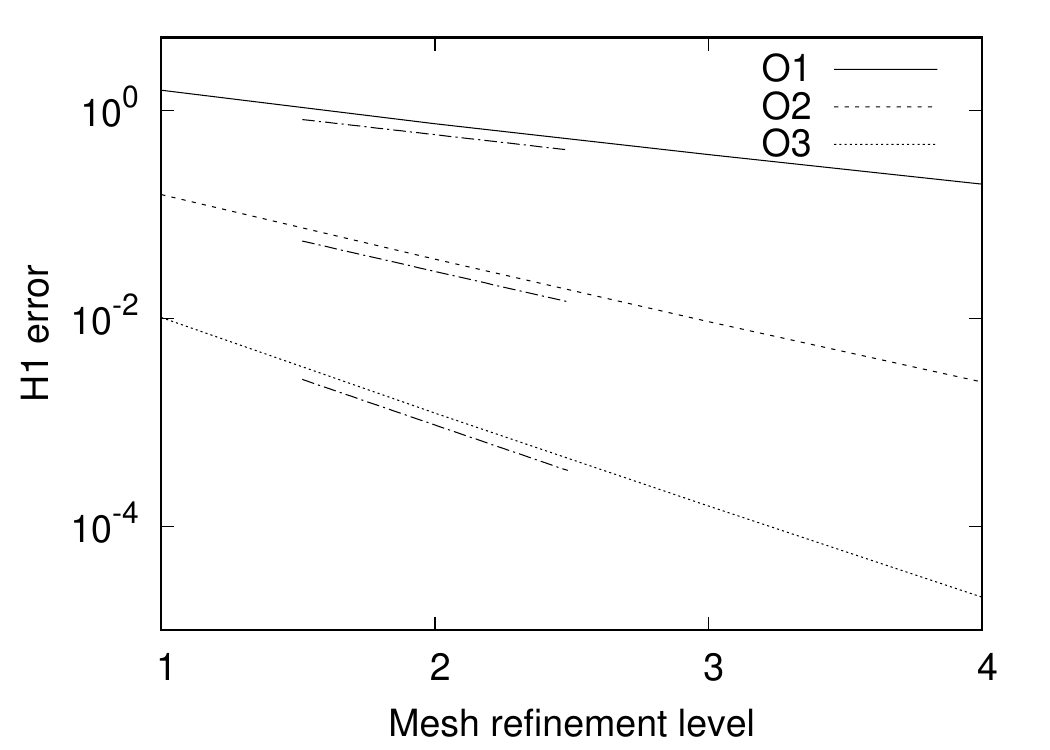}
   \end{center}
   \caption{ The left figure shows an elastically deformed Gyroid structure with a displacement field obtained through CutFEM solution on a regular grid with a predefined force function, and the right figure shows the $H_1$ convergence plot of a regularized CutFEM solution for linear, quadratic, and cubic Lagrangian elements.}
   \label{fig_cutfem}
\end{figure}

MFEM provides two alternative approaches for constructing integration rules over discretely prescribed cut surfaces and volumes.
Both methods enable the computation of
\[
 S = \int_{\phi = 0} u(x) \, ds, \quad
 V = \int_{\phi > 0} u(x) \, dx,
\]
where the level set function $\phi$ specifies the cut, and $u$ is the integrand.
The first approach relies on the external Algoim package \cite{Saye2014}, which
works for quadrilateral and hexahedral elements by performing dimension reduction to construct a
combination of 1D quadrature rules.
The second approach uses the moment-fitting algorithm introduced in \cite{Muller2013}.
This method constructs a set of basis functions for each element to define and solve a local under-determined system for the vector of quadrature weights.
All surface and volume integrals required to form the system are reduced to 1D integration over intersected segments.
The newly developed integration techniques allow for fast and efficient implementation of high-order CutFEM \cite{Burman2015} discretizations as demonstrated in \Cref{fig_cutfem}.

\subsubsection{Shifted boundary method.}
\begin{figure}
   \includegraphics[width=0.8\linewidth]{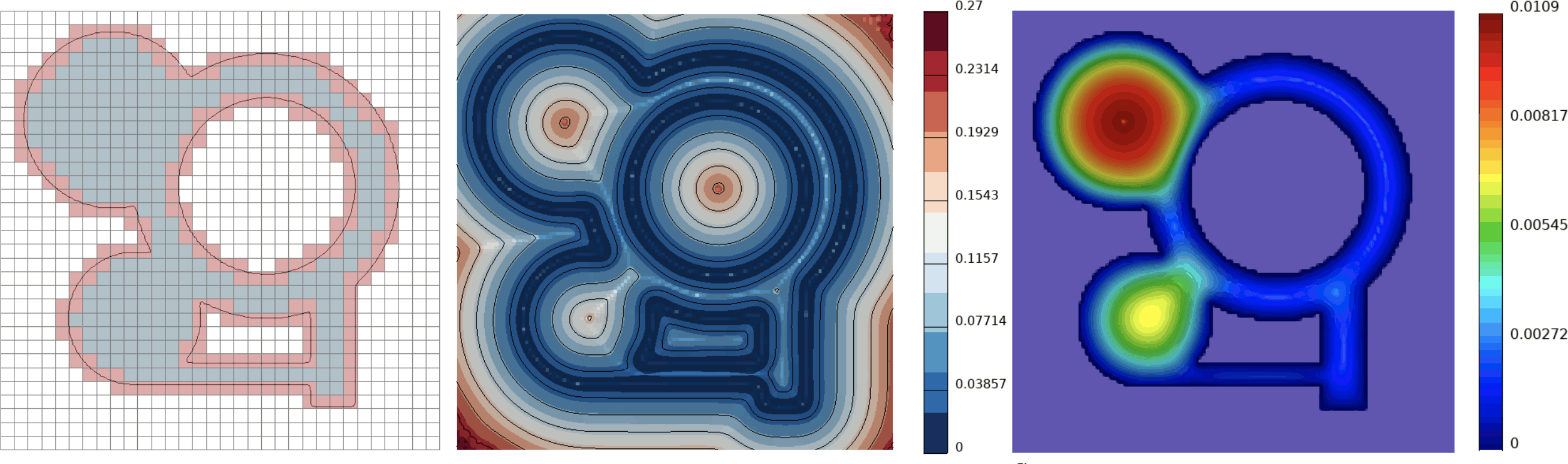}
   \caption{Example shifted boundary calculation.
   Left: true boundary with background mesh and surrogate domain.
   Middle: discrete distance field.
   Right: SBM solution.}
   \label{fig_shift}
\end{figure}

MFEM also provides an implementation of the Shifted Boundary Method (SBM)
\cite{Atallah2020}, which is a technique that avoids integration in cut
elements by using a surrogate computational domain.
The method uses a distance function to the true boundary to enforce
the required boundary conditions on the (non-aligned) surrogate mesh faces,
thereby \textit{shifting} the location where the boundary conditions are imposed,
see \Cref{fig_shift}.
The enforcement is performed weakly, through face integrals on the surrogate faces.
The desired accuracy is obtained by combining the distance function with a Taylor expansion of certain order.
These techniques are illustrated in MFEM's \textit{Shifted} miniapp.

\subsubsection{Distance and extrapolation solvers.}
Discrete distance solvers are often needed for computations that involve level set functions.
MFEM's \textit{Distance} miniapp
demonstrates the capability to compute the shortest path
through the computational mesh
to a given point source or to the zero level set of a given function (see \Cref{fig_shift}).
The implemented methods include the heat method described in \cite{Crane2017},
and the $p$-Laplacian method from \cite{Belyaev2015}.

Extrapolation is another useful capability in immersed frameworks.
The \textit{Extrapolation} miniapp extrapolates a finite element grid function
from known values on a set of elements to the rest of the domain.
The miniapp supports the PDE-based approaches from \cite{Aslam2004} and
\cite{Bochkov2020}, both of which rely on solving a sequence of advection
problems in the direction of the unknown parts of the domain.
The extrapolation can be up to third-order in a
limited band around the zero level set.
 
\subsection{Partial assembly on NURBS patches}
In addition to the existing support for partial assembly on elements, MFEM now supports partial assembly on NURBS patches in the settings of isogeometric analysis.
Significant computational savings are possible when assembling patch-wise rather than element-wise, as the tensor-product structure exists on the entire patch of elements, allowing for more benefit in sum factorization.
In this sense, a NURBS patch, containing many elements, can be considered as a very high-order single element.
This new feature is demonstrated in the \textit{NURBS} miniapp.

Optimization of patch-wise assembly has been done in works such as \cite{Hiemstra2019}.
At the patch level, 1D basis functions in the sum factorization may span multiple elements, but computations for a pair of test and trial basis functions are limited to the intersection of their spans.
We further optimize by computing a reduced quadrature rule for each basis function.
Given the smoothness of patch basis functions, quadrature rules chosen for a particular order on one element may have more points than necessary to achieve accuracy, so a reduced rule is possible.
We compute reduced rules automatically by the algebraic non-negative least squares (NNLS) method \cite{Lawson1995}, rather than deriving and implementing reduced rules for every possible case.
These reduced rules are computed on the fly, for the choice of NURBS space, patch, and bilinear form, and stored for use throughout a system solve or simulation.

The newly introduced NNLS solver is also a generally useful addition, which in particular could be used for computing reduced quadrature rules in other contexts.
 
\subsection{Automatic differentiation (\texorpdfstring{$\partial$}{∂}FEM)}
In recent years, there has been an increased demand for automatic differentiation (AD) of numerical simulations, due in part to the rise of machine learning (ML) frameworks (see e.g.~\cite{abadi2016tensorflow,paszke2017automatic}).
AD includes a set of well-established techniques for evaluating derivatives of functions written as computer programs with traditional implementation approaches based on operator overloading~\cite{griewank1996algorithm} or source transformation~\cite{TapenadeRef13}.
Some of these conventional approaches require the use of a domain-specific language~\cite{paszke2017automatic}, limiting the portability and the number of possible hardware and software execution platforms.
On the other hand,  MFEM is used in an HPC context for complex applications composed of libraries in different languages; traditional AD approaches would necessitate significant changes to the data types and code organization, making their application to MFEM more challenging.

To address this, a collaborative effort, the $\partial \mbox{FEM}$ project targeting the integration of the Enzyme tool~\cite{moses_reverse-mode_2021} and developing further a native MFEM AD implementation based on operator overloading and templating, emerged from the need to automatically compute residuals, Jacobians, and Hessians in FEM discretizations and optimal design applications.
In contrast to traditional tools, Enzyme performs forward and reverse-mode differentiation on the LLVM compiler intermediate representation~\cite{LLVM:CGO04}, enabling it to synthesize fast derivatives of programs in any language whose compiler targets LLVM, as well as a wide variety of parallel frameworks and hardware architectures, all in a single tool.

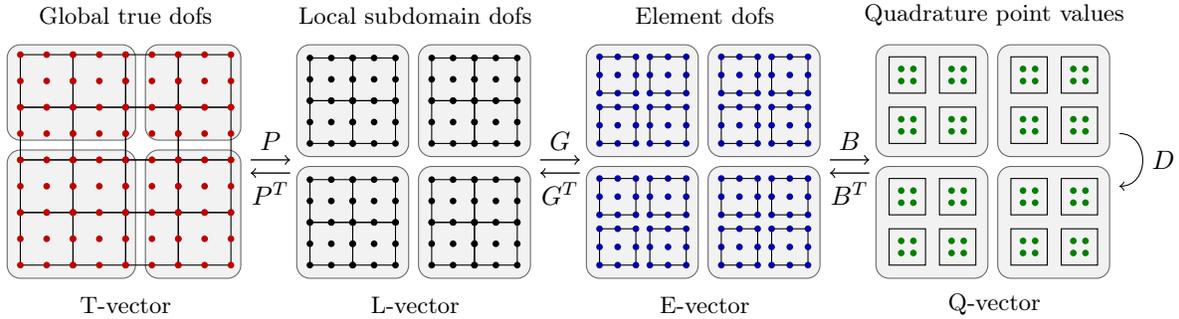
\begin{figure}
    \centering\newcommand{\RoundRect}[4]{\draw[
    rounded corners=5,
    black!60!white,
    fill=black!5!white,
  ] (#1,#2) rectangle ++(#3,#4);
}

\newcommand{\DofSquare}[5]{\draw[black] (#1,#2) rectangle ++(#3,#4);
  \draw node[fill,circle,inner sep=0pt,minimum size=2.5pt,#5] at (#1, #2) {};
  \draw node[fill,circle,inner sep=0pt,minimum size=2.5pt,#5] at ({#1+#3/2}, #2) {};
  \draw node[fill,circle,inner sep=0pt,minimum size=2.5pt,#5] at ({#1+#3}, #2) {};
  \draw node[fill,circle,inner sep=0pt,minimum size=2.5pt,#5] at (#1, {#2+#4/2}) {};
  \draw node[fill,circle,inner sep=0pt,minimum size=2.5pt,#5] at ({#1+#3/2}, {#2+#4/2}) {};
  \draw node[fill,circle,inner sep=0pt,minimum size=2.5pt,#5] at ({#1+#3}, {#2+#4/2}) {};
  \draw node[fill,circle,inner sep=0pt,minimum size=2.5pt,#5] at (#1, {#2+#4}) {};
  \draw node[fill,circle,inner sep=0pt,minimum size=2.5pt,#5] at ({#1+#3/2}, {#2+#4}) {};
  \draw node[fill,circle,inner sep=0pt,minimum size=2.5pt,#5] at ({#1+#3}, {#2+#4}) {};
}

\newcommand{\QuadSquare}[5]{\draw[black] (#1,#2) rectangle ++(#3,#4);
  \draw node[fill,circle,inner sep=0pt,minimum size=2.5pt, #5] at (#1+#3/3, #2+#4/3) {};
  \draw node[fill,circle,inner sep=0pt,minimum size=2.5pt, #5] at (#1+2*#3/3, #2+#4/3) {};
  \draw node[fill,circle,inner sep=0pt,minimum size=2.5pt, #5] at (#1+#3/3, #2+2*#4/3) {};
  \draw node[fill,circle,inner sep=0pt,minimum size=2.5pt, #5] at (#1+2*#3/3, #2+2*#4/3) {};
}

\newcommand{\DofGrid}[5]{\DofSquare{#1}{#2}{#3}{#4}{#5}
  \DofSquare{#1+#3}{#2}{#3}{#4}{#5}
  \DofSquare{#1}{#2+#4}{#3}{#4}{#5}
  \DofSquare{#1+#3}{#2+#4}{#3}{#4}{#5}
}

\def\spacingfactor{0.85}

\newcommand{\DofElem}[5]{\DofSquare{#1}{#2}{#3*\spacingfactor}{#4*\spacingfactor}{#5}
  \DofSquare{#1+#3/\spacingfactor}{#2}{#3*\spacingfactor}{#4*\spacingfactor}{#5}
  \DofSquare{#1}{#2+#4/\spacingfactor}{#3*\spacingfactor}{#4*\spacingfactor}{#5}
  \DofSquare{#1+#3/\spacingfactor}{#2+#4/\spacingfactor}{#3*\spacingfactor}{#4*\spacingfactor}{#5}
}

\newcommand{\QuadElem}[5]{\QuadSquare{#1}{#2}{#3*\spacingfactor}{#4*\spacingfactor}{#5}
  \QuadSquare{#1+#3/\spacingfactor}{#2}{#3*\spacingfactor}{#4*\spacingfactor}{#5}
  \QuadSquare{#1}{#2+#4/\spacingfactor}{#3*\spacingfactor}{#4*\spacingfactor}{#5}
  \QuadSquare{#1+#3/\spacingfactor}{#2+#4/\spacingfactor}{#3*\spacingfactor}{#4*\spacingfactor}{#5}
}

\newcommand{\RoundRectGrid}[2]{\RoundRect{0}{0}{#1}{#2}
  \RoundRect{#1 + 0.075}{0}{#1}{#2}
  \RoundRect{0}{#2 + 0.075}{#1}{#2}
  \RoundRect{#1 + 0.075}{#2 + 0.075}{#1}{#2}
}

\begin{tikzpicture}[scale=1.75]

  \RoundRect{0}{0}{0.975}{0.975}
  \RoundRect{1.05}{0}{0.725}{0.975}
  \RoundRect{0}{1.05}{0.975}{0.725}
  \RoundRect{1.05}{1.05}{0.725}{0.725}

  \DofGrid{0.1}{0.1}{0.4}{0.4}{red!75!black}
  \DofGrid{0.9}{0.1}{0.4}{0.4}{red!75!black}
  \DofGrid{0.1}{0.9}{0.4}{0.4}{red!75!black}
  \DofGrid{0.9}{0.9}{0.4}{0.4}{red!75!black}

  \draw[->, line width=0.5pt] (1.85, 0.9) -- node[midway,above] {$P$} ++(.3,0.0);
  \draw[<-, line width=0.5pt] (1.85, 0.8) -- node[midway,below] {$P^T$} ++(.3,0.0);
  \node[align=center] at (0.9, -0.2) {\small T-vector};
  \node[align=center] at (0.9, 2) {\small Global true dofs};

  \begin{scope}[shift={(2.2,0)}]
    \RoundRectGrid{0.85}{0.85}

    \DofGrid{0.1}{0.1}{0.325}{0.325}{black}
    \DofGrid{1.025}{0.1}{0.325}{0.325}{black}
    \DofGrid{0.1}{1.025}{0.325}{0.325}{black}
    \DofGrid{1.025}{1.025}{0.325}{0.325}{black}

    \draw[->, line width=0.5pt] (1.85, 0.9) -- node[midway,above] {$G$} ++(.3,0.0);
    \draw[<-, line width=0.5pt] (1.85, 0.8) -- node[midway,below] {$G^T$} ++(.3,0.0);
    \node[align=center] at (0.9, -0.2) {\small L-vector};
    \node[align=center] at (0.9, 2) {\small Local subdomain dofs};
  \end{scope}

  \begin{scope}[shift={(4.4,0)}]
    \RoundRectGrid{0.85}{0.85}

    \DofElem{0.1}{0.1}{0.325}{0.325}{blue!65!black}
    \DofElem{1.025}{0.1}{0.325}{0.325}{blue!65!black}
    \DofElem{0.1}{1.025}{0.325}{0.325}{blue!65!black}
    \DofElem{1.025}{1.025}{0.325}{0.325}{blue!65!black}

    \draw[->, line width=0.5pt] (1.85, 0.9) -- node[midway,above] {$B$} ++(.3,0.0);
    \draw[<-, line width=0.5pt] (1.85, 0.8) -- node[midway,below] {$B^T$} ++(.3,0.0);
    \node[align=center] at (0.9, -0.2) {\small E-vector};
    \node[align=center] at (0.9, 2) {\small Element dofs};
  \end{scope}

  \begin{scope}[shift={(6.6,0)}]
    \RoundRectGrid{0.85}{0.85}

    \QuadElem{0.1}{0.1}{0.325}{0.325}{green!50!black}
    \QuadElem{1.025}{0.1}{0.325}{0.325}{green!50!black}
    \QuadElem{0.1}{1.025}{0.325}{0.325}{green!50!black}
    \QuadElem{1.025}{1.025}{0.325}{0.325}{green!50!black}

    \draw [->] (1.85, 1.1) to[out=0,in=0,looseness=1.5] node[midway,right] {$D$} (1.85, 0.7) ;
    \node[align=center] at (0.9, -0.2) {\small Q-vector};
    \node[align=center] at (0.9, 2) {\small Quadrature point values};
  \end{scope}

\end{tikzpicture}     \caption{Finite element operators, $A_p$, have a natural decomposition, $A_p = P^T G^T B^T D B G P$, which exposes multi-level parallelism and allows for AD-friendly, matrix-free, memory-efficient implementations that assemble and store only the innermost, pointwise operator component \cite{Anderson2020}.}
    \label{fig_feod}
\end{figure}

MFEM's infrastructure is built around the finite element operator decomposition, shown in~\Cref{fig_feod}, which encapsulates a generic description of an assembly procedure in a finite element library.
Instead of following traditional approaches that apply automatic differentiation at a global level and treat the implementation as a black box, the operator decomposition allows MFEM to treat derivatives on the innermost level at the quadrature points ($D$).
The operators transferring data from a global level to subdomain, element, and quadrature levels, i.e., $P$, $G$, and $B$, are linear and topological: they do not depend on the solution, physical coordinates or design parameters.
Therefore, they can be excluded from the differentiation loop, saving a significant amount of memory and computations.
The decomposition confines the code modifications to the integration point level, allowing full automation of the discretization process for complex non-linear problems.

While the $\partial \mbox{FEM}$ effort is still a work-in-progress, a proof of concept can be found in the \textit{Hooke} miniapp, which implements a finite-strain elasticity discretization with arbitrarily user-defined material models and automatically generated Jacobians (derivatives of the residual equation with respect to the displacement).
The derivatives in this example can either be generated by leveraging Enzyme or MFEM's internal dual number type implementation.
 
\section{Meshing}
MFEM includes advanced meshing support such as general element types, conforming
and non-conforming adaptive mesh refinement, high-order meshing and more, see
\cite{Anderson2020}. In this section we review recent progress in MFEM's mesh
optimization algorithms and describe the addition of submesh capabilities for
multi-domain problems.

\subsection{High-order mesh optimization}
The high-order mesh optimization framework in MFEM is based on
the Target-Matrix Optimization Paradigm (TMOP).
This framework enables precise control over local
mesh quality, inferred through the local Jacobian $A_{d\times d}$ of
the transformation from the reference to physical space coordinates,
while still optimizing the mesh globally.

The first step with TMOP is to define a target transformation matrix
$W_{d\times d}$ analogous to $A$.
Construction of $W$ is typically driven by the fact that any Jacobian matrix can be
composed as a function of four fundamental geometric properties:
\begin{equation}
\label{eq_W}
W_{d\times d} = \underbrace{\zeta}_{\text{[size]}} \underbrace{R_{d\times d}}_{\text{[rotation]}}
\underbrace{Q_{d\times d}}_{\text{[skewness]}} \underbrace{D_{d\times d}}_{\text{[aspect ratio]}}.
\end{equation}
namely size/volume, rotation, skewness, and aspect-ratio.  In general, each of
the components of $W$ in \eqref{eq_W} can vary spatially based on the desired
properties at each point in the mesh.  A detailed description of
target construction procedure is provided in \cite{knupp2019target}.

With the target transformation $W$ defined, a mesh quality metric, $\mu(T)$, $T=AW^{-1}$,
is used to measure the difference between the transformations $A$ and $W$.
Mesh quality metrics are categorized based on the geometric parameters they depend on. For example,
shape metrics $\mu_{Sh}$ depend on the element skewness and
aspect-ratio, size metrics $\mu_{Sz}$ depend on the element size.
There are also composite
metrics that depend on some combination of the four parameters (e.g., $\mu_{ShSz}$ and $\mu_{ShSzOr}$).
In practice, the user defines $W$ and chooses a mesh quality metric based on the geometric
properties they wish to optimize.

The quality metric $\mu(T)$ is used to define the TMOP objective function
\begin{equation}
\label{eq_F_full}
  F(\bm{x}) = \sum_{E \in \mathcal{M}} \int_{E_t} \mu(T(\bm{x})) \, d\bm{x}_t,
\end{equation}
where $E_t$ is the target element determined by $W$, and $d\bm{x}_t$ denotes
integration over the target element.
Optimal node locations are determined using Newton's method by solving
$\partial F(\bm{x}) / \partial \bm{x} = 0$.
Note that since \eqref{eq_F_full} can be represented solely in terms of finite element operations,
it extends to all element types (quadrilaterals, hexahedra, simplices, pyramids) in 2D and 3D, supports
$h$/$p$\,-refined meshes, and can leverage advances in finite element assembly technique targeting
GPU acceleration.

\begin{figure}
  \centering
  \includegraphics[width=0.2\linewidth]{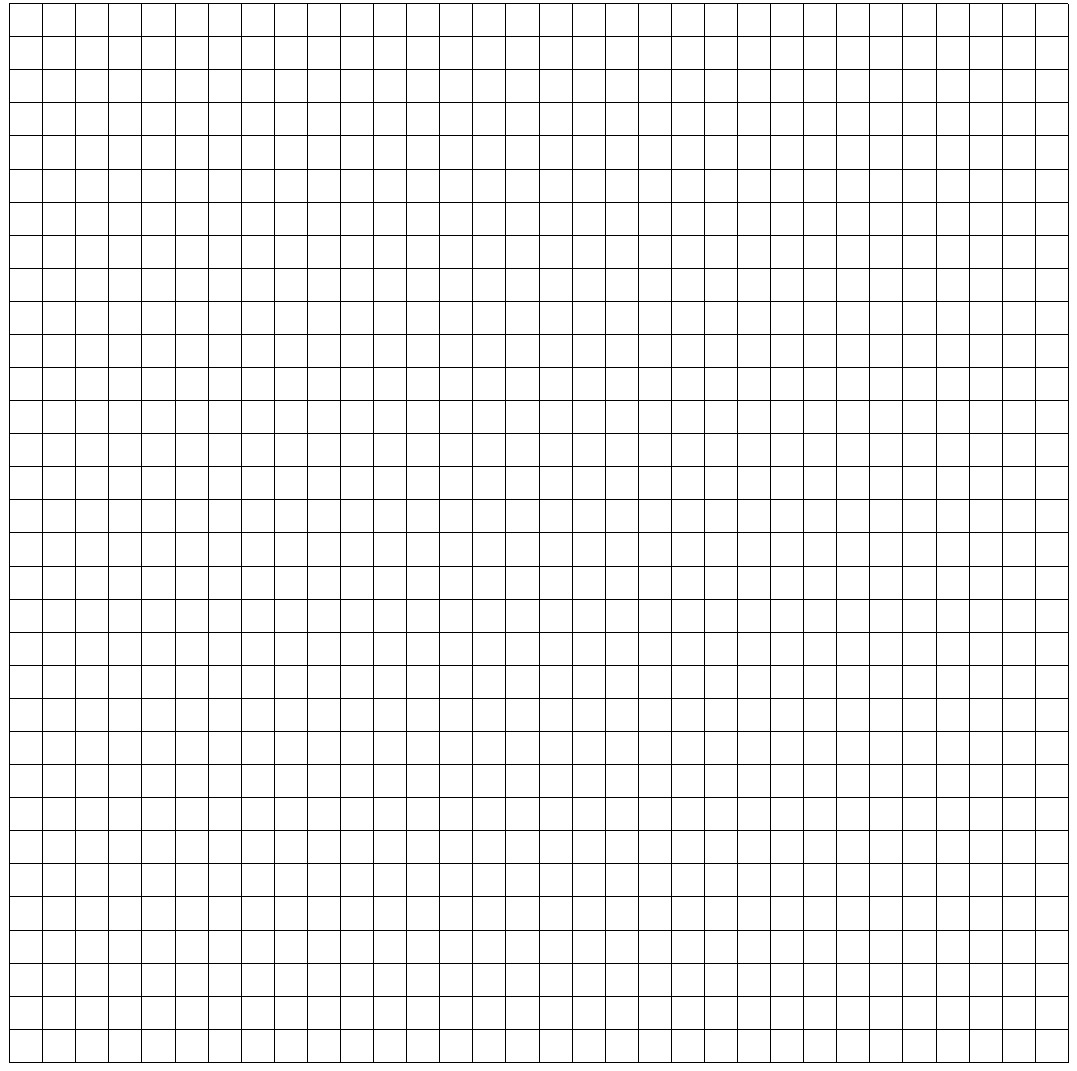}\includegraphics[width=0.2\linewidth]{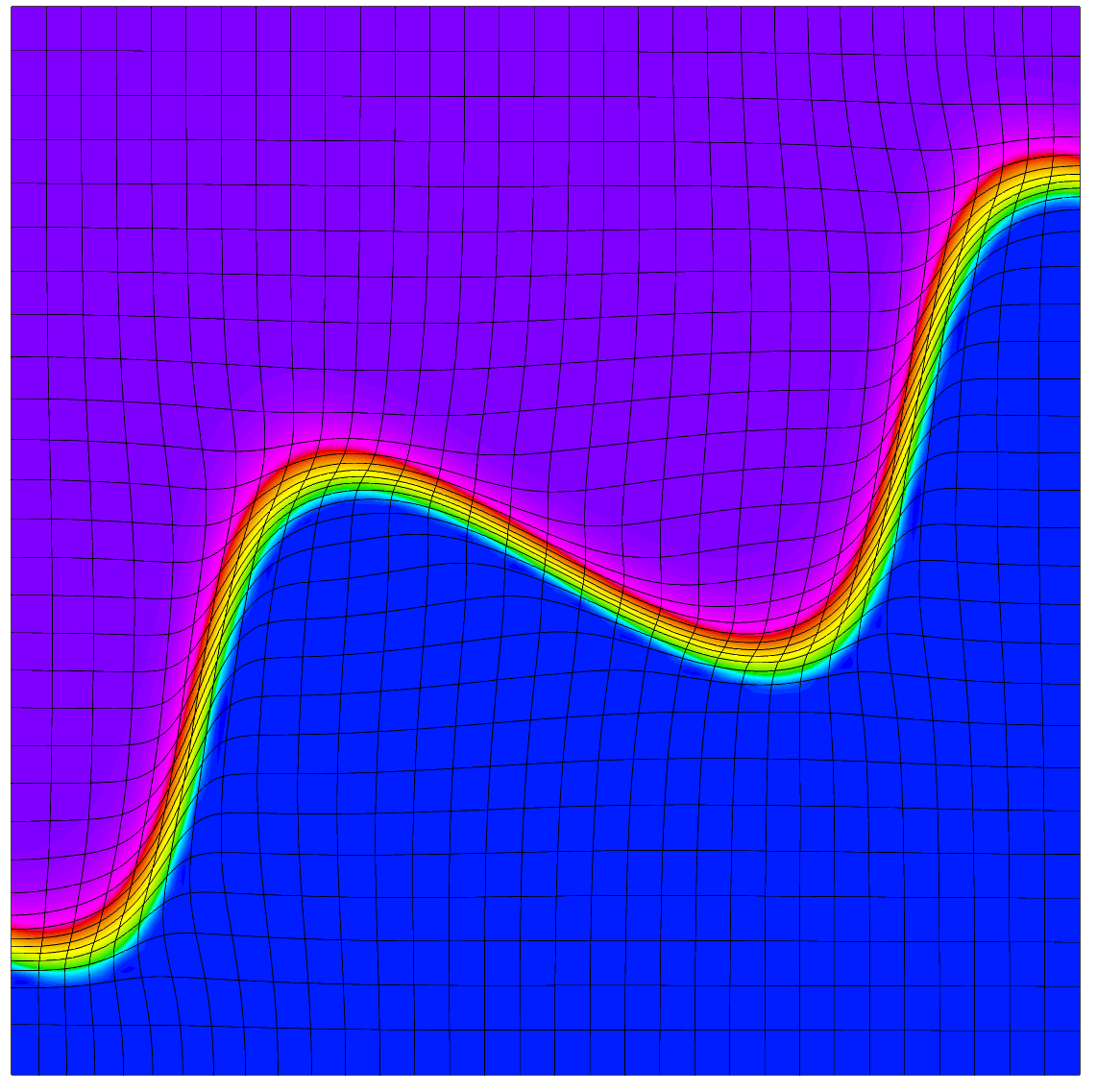}\includegraphics[width=0.2\linewidth]{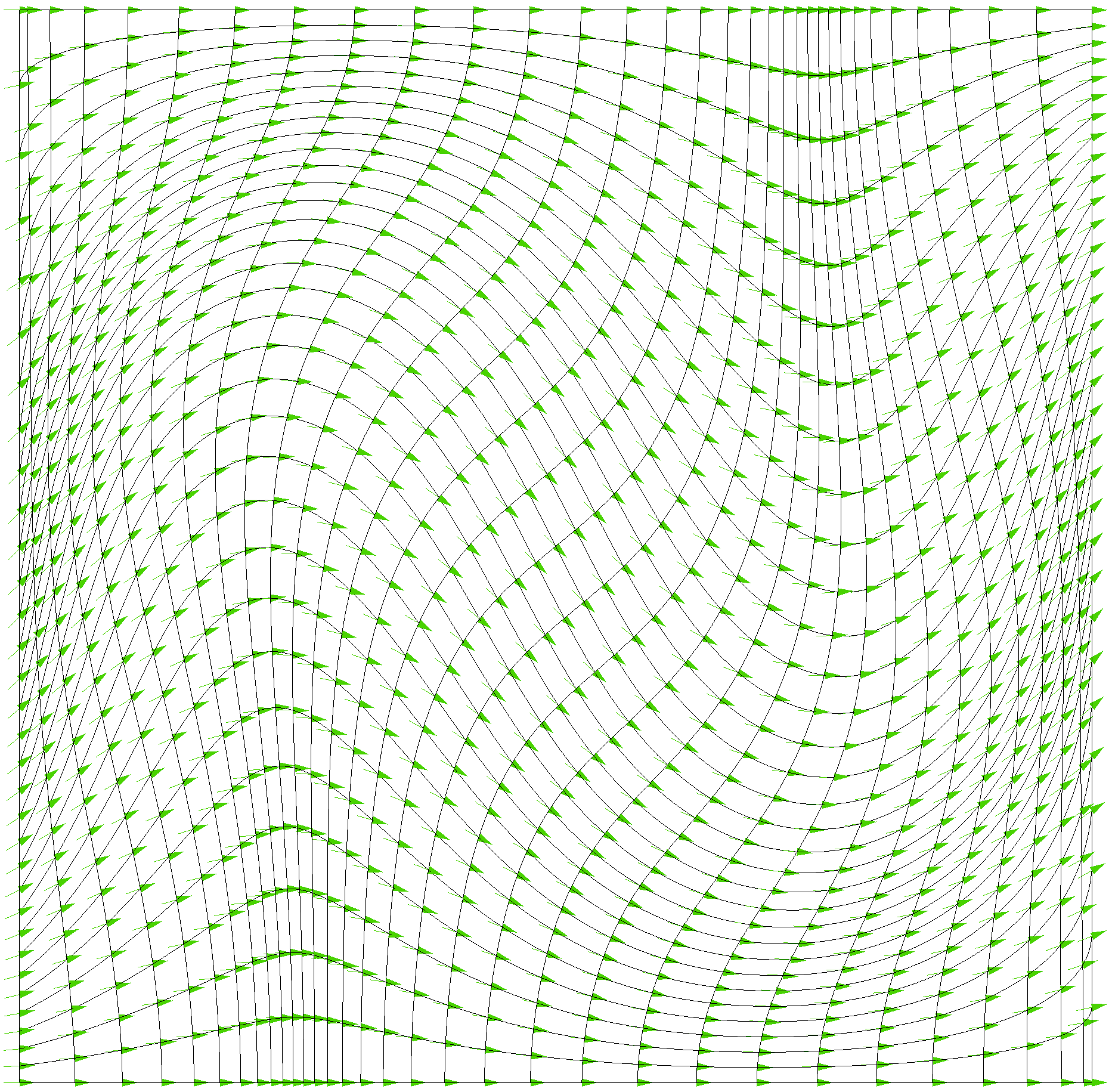}
  \caption{
    Simulation-driven adaptivity to morph a uniform mesh (left) for shape and size adaptivity using a scalar solution (center) and size and orientation adaptivity using a vector solution (right).
  }
\label{fig_tmop_sim_tar}
\end{figure}

The effectiveness of TMOP for mesh quality improvement
largely depends on the prescribed target and the mesh quality metric, and we have
advanced the state-of-the-art in each of these areas.
We have developed approaches for automated target construction for different mesh adaptivity
goals \cite{knupp2019target,Dobrev2020,Dobrev2021}. For general mesh quality
improvement, the target geometric properties are set based on the ideal element
type (e.g., cube or regular tetrahedron).
For simulation-driven optimization, the discrete solution or a derived quantity
(e.g., error estimate) is used to set the target geometric properties.
\Cref{fig_tmop_sim_tar} shows an example of simulation-driven adaptivity
where a uniform mesh is morphed to adapt its shape and size with respect to a
scalar solution and its size and orientation for a vector solution.

\begin{figure}
\centering
\includegraphics[width=0.5\linewidth]{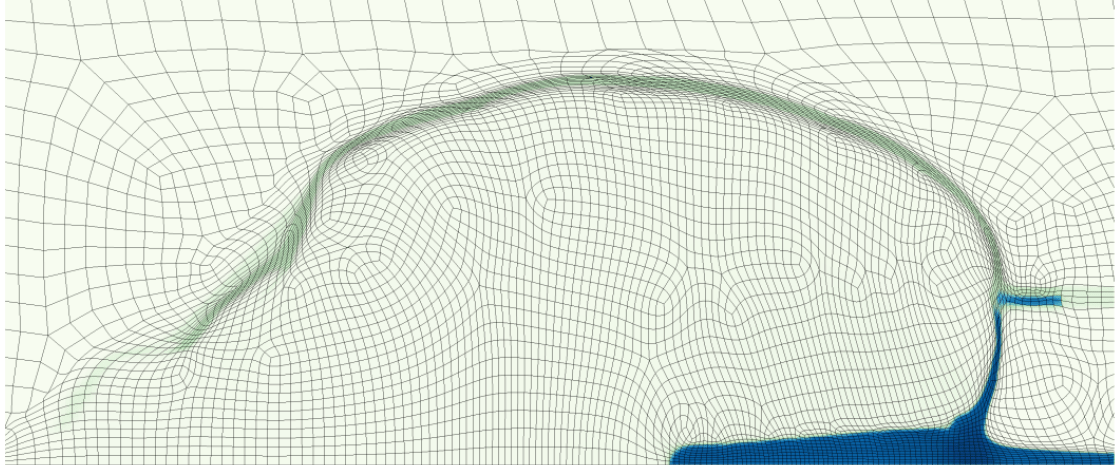}
\caption{
  Shape and size optimization for a 2D shaped charge simulation using
an asymptotically balanced compound metric, $\mu_{ShSz} = \mu_{Sh}+\frac{3}{2}\mu_{Sz}$ \cite{TMOP2024COMBO}.
}
\label{fig_tmop_combo_metric}
\end{figure}

We have also analyzed the theoretical properties of different mesh quality
metrics in recent work \cite{Knupp2020,Knupp2023} to identify well
posed metrics of different types.
Compound metrics of the form $\mu_{ShSz}=\mu_{Sh}+\gamma\mu_{Sz}$, $\gamma \in
\mathbb{R}^+$ are effective for simulation-driven optimization, but require
tuning $\gamma$ on a case-by-case basis as its subcomponents may scale
differently with mesh distortion.  We have studied the asymptotic properties of
different compound metrics to address this issue. We have also developed new
asymptotically balanced compound metrics of the form
$\mu_{ShSz}=\mu_{Sh}^{\alpha}+\lambda\mu_{Sz}^{\beta}$, $\{\alpha,\beta\} \in
\mathbb{R}^+$, that are balanced for a prescribed range of $\lambda$ \cite{TMOP2024COMBO}. These compound metrics have been impactful for shape and size
adaptivity in arbitrary Lagrangian-Eulerian (ALE) simulations (cf.~\cite{Anderson2018,
Rieben2020}), without requiring user intervention; see
\Cref{fig_tmop_combo_metric} for an example.  Finally, we have also developed
new mesh quality metrics that can simultaneously untangle the mesh and improve
worst element quality. These metrics are crucial for problems with severe or
localized mesh distortion.

\begin{figure}
  \newdimen\imageheight
  \centering
  \settoheight{\imageheight}{\includegraphics[width=0.2\linewidth]{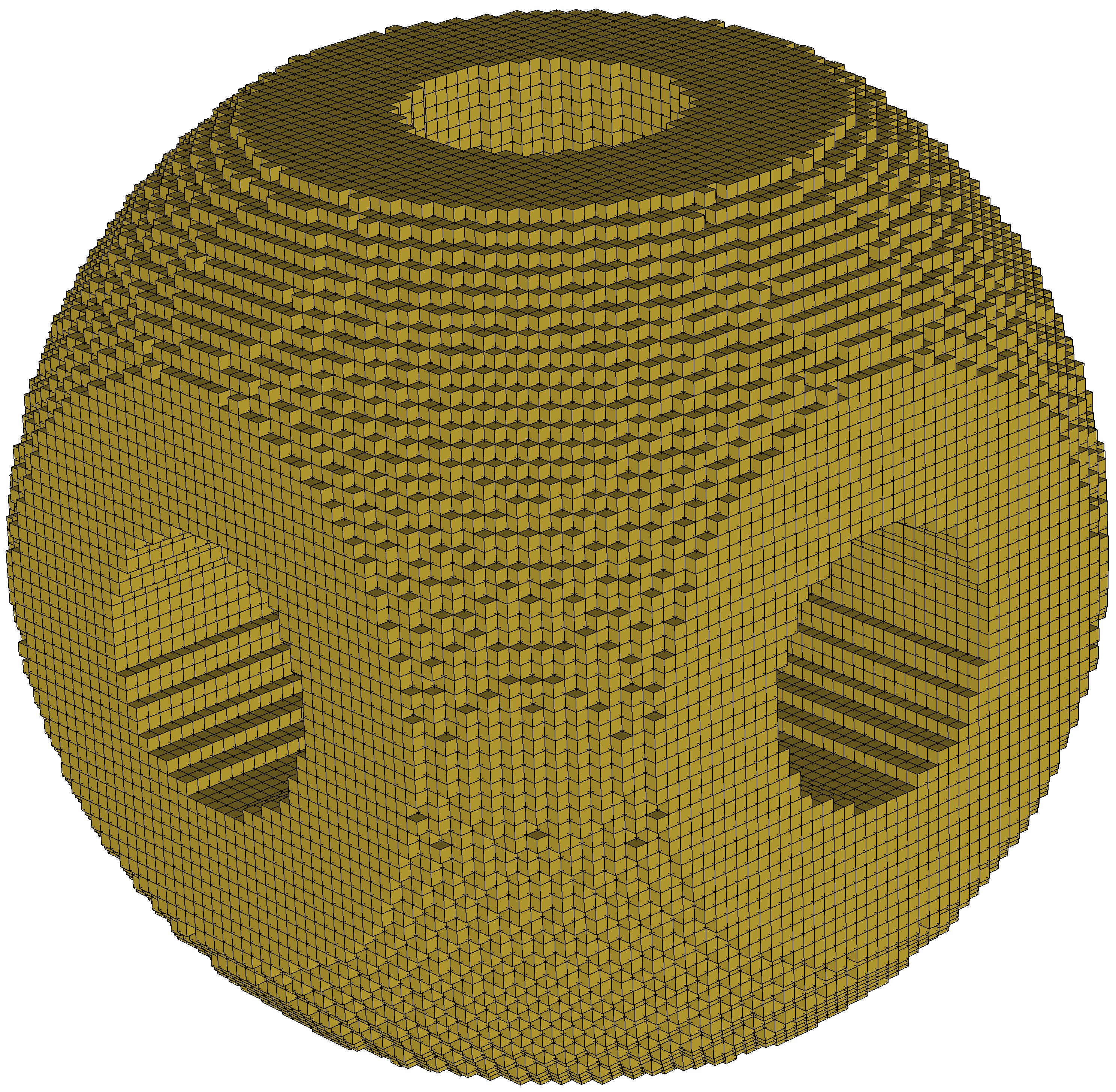}}\includegraphics[height=\imageheight]{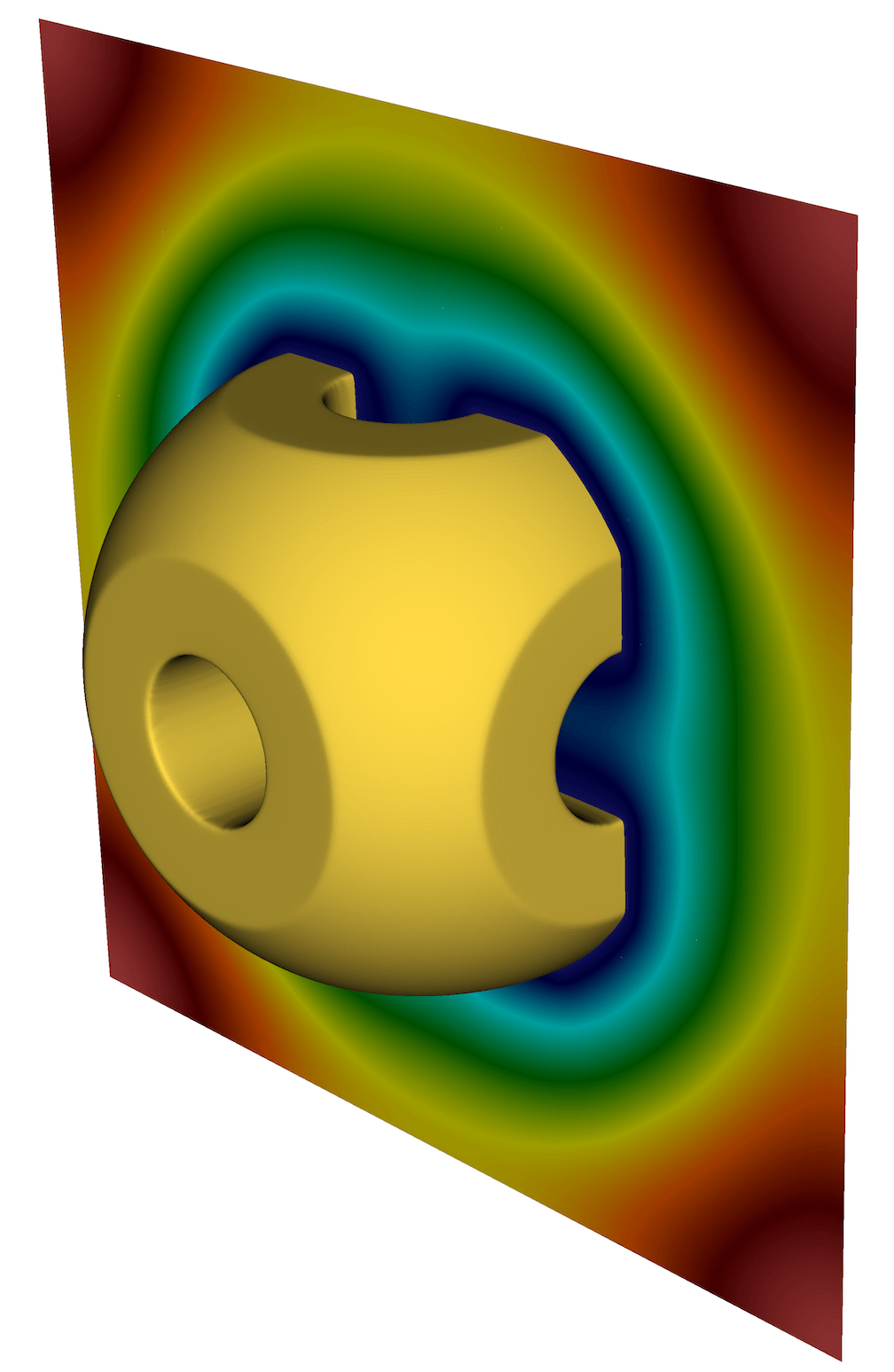}\includegraphics[width=0.2\linewidth]{fig/tmop_cgs_3d_origmesh}\includegraphics[width=0.2\linewidth]{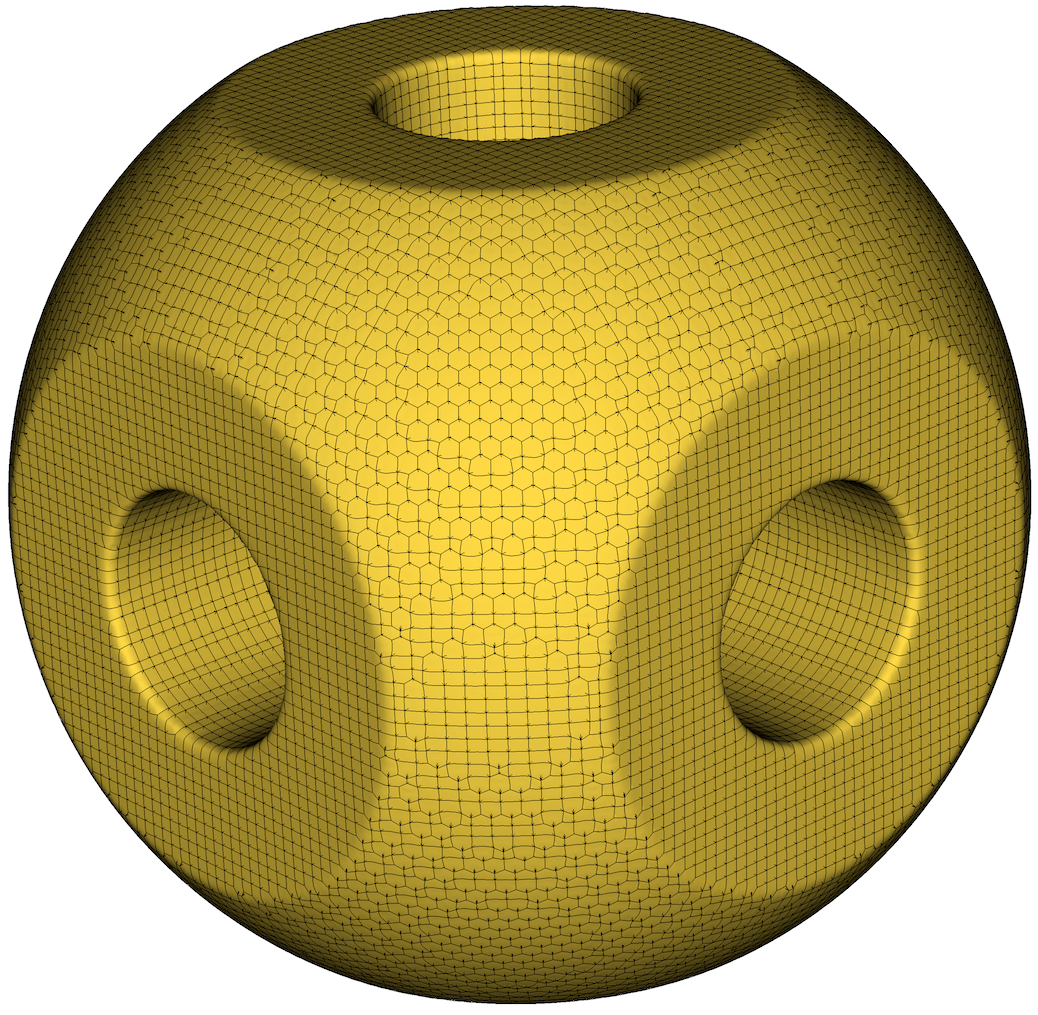}
  \caption{
    Discrete function along with its zero isosurface representing a 3D curvilinear surface (left).
    Initial uniform hex mesh (center) morphed to align with the target surface (right).
  }
  \label{fig_tmop_fit_3d}
\end{figure}

\begin{figure}
\includegraphics[width=0.5\linewidth]{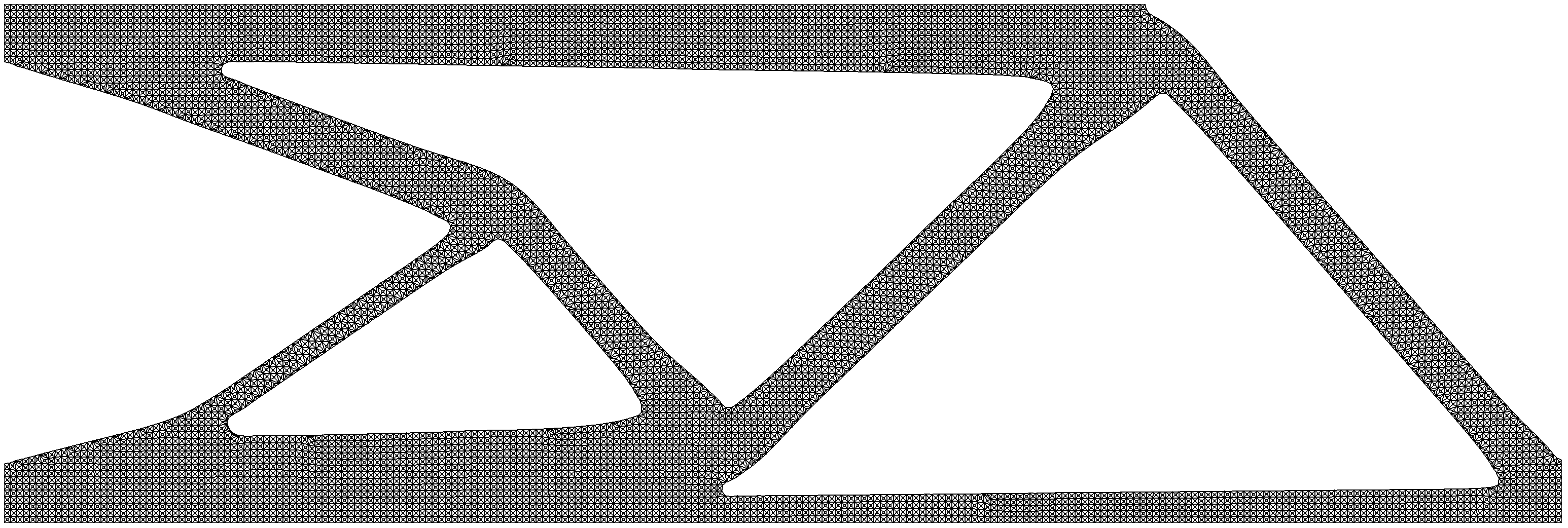}\includegraphics[width=0.2\linewidth]{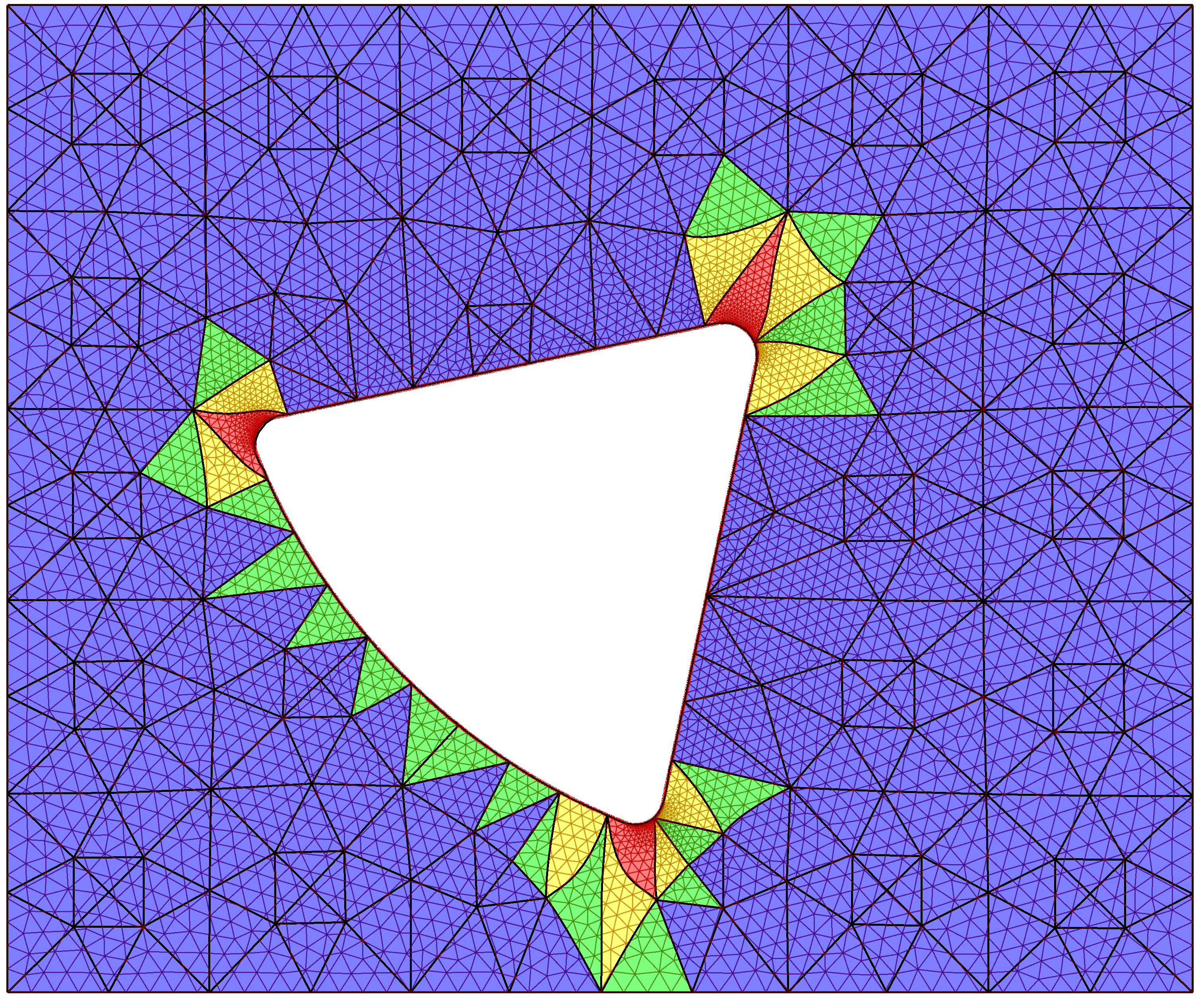}
\caption{
  Left panel: topology-optimized cantilever beam where the surface fitting formulation is used to obtain a body-fitted mesh at each TO iteration.
  Right panel: $rp$-refined body-fitted mesh for a 2D version of the Apollo capsule. Different colors represent different polynomial orders.}
\label{fig_tmop_fit}
\end{figure}

In recent work, we have augmented the TMOP-based formulation
with a penalization term to weakly enforce alignment of a selected set of mesh
nodes with an implicit surface, while simultaneously optimizing the mesh
quality \cite{Barrera2023}. This formulation requires that the target surface be prescribed as the
zero isocontour of a smooth discrete level-set function, which is often the
case for representing interfaces in multimaterial configurations and
evolving geometries in shape and topology optimization. The modified objective function is
\begin{equation}
\label{eq_F_full_sigma}
  F(\bm{x}) = \underbrace{\sum_{E \in \mathcal{M}} \int_{E_t}
  \mu(T(\bm{x})) d\bm{x}_t}_{F_{\mu}} +
  \underbrace{w_{\sigma} \sum_{s \in S} \sigma^2(\bm{x}_s)}_{F_{\sigma}}.
\end{equation}
Here, $F_{\mu}$ is the mesh quality term, and
$F_{\sigma}$ is a penalty term that depends on the penalization weight
$w_{\sigma}$, the set of mesh nodes $\mathcal{S}$ to be aligned to the level set, and
the level set function $\sigma(\bm{x})$ evaluated at the positions
$\bm{x}_s$ of all nodes $s \in \mathcal{S}$.
This formulation has proven to be robust at obtaining body-fitted meshes for
complex curvilinear domains, and is helping circumvent the bottleneck of
high-order mesh generation; see \Cref{fig_tmop_fit_3d,fig_tmop_fit}. We have further augmented this formulation with MFEM's
$p$-adaptivity framework to obtain mixed-order meshes, such that high-order
elements are only used adjacent to regions of high-curvature in the target
surface. This $rp$-adaptivity approach has proven to be robust at obtaining
mixed-order meshes with same geometric accuracy at a lower computational cost
in comparison to a constant-order mesh \cite{mittal2024mixed}.

\begin{figure}
  \raisebox{-0.5\height}{\includegraphics[width=0.2\linewidth]{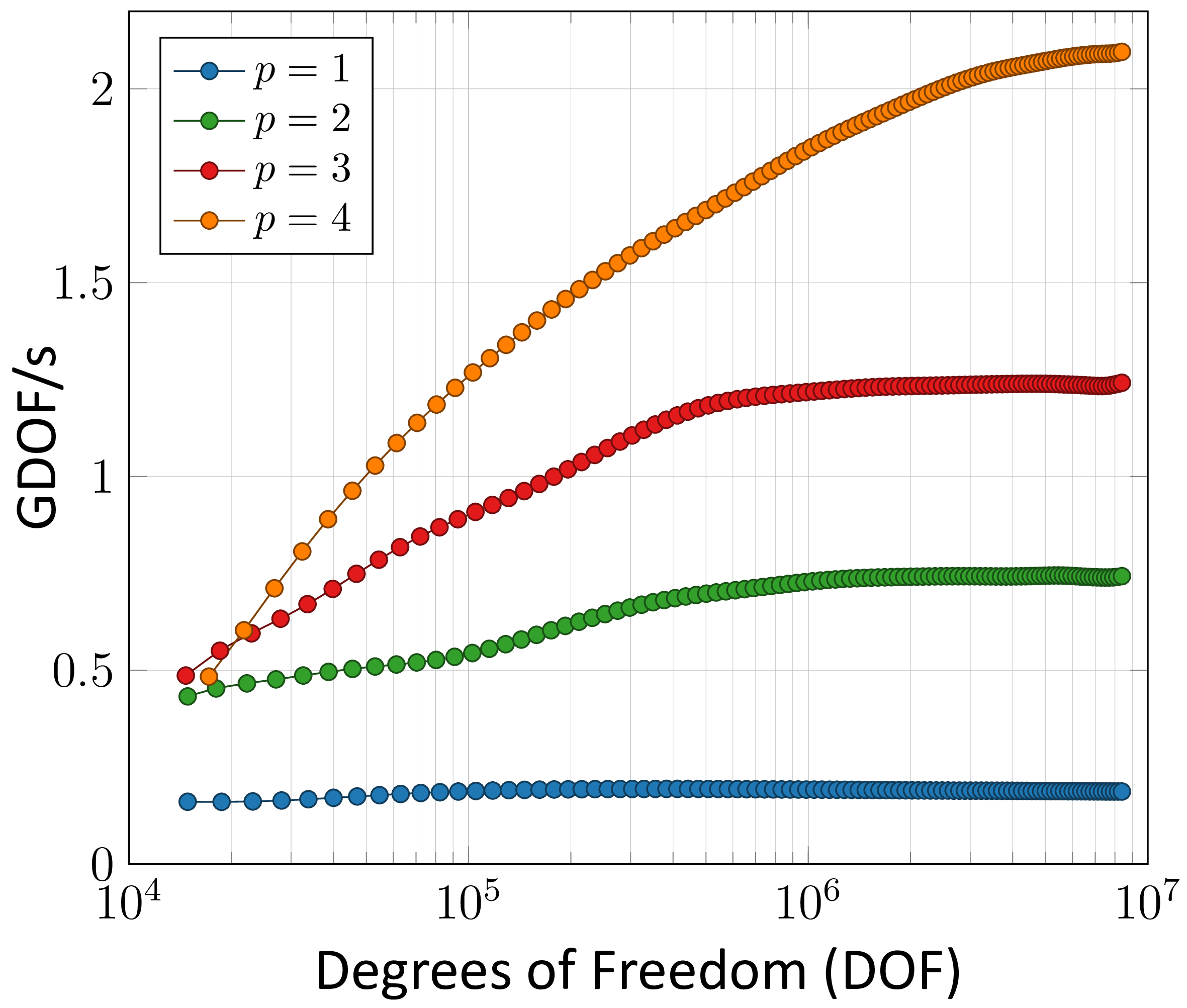}}\raisebox{-0.5\height}{\includegraphics[width=0.2\linewidth]{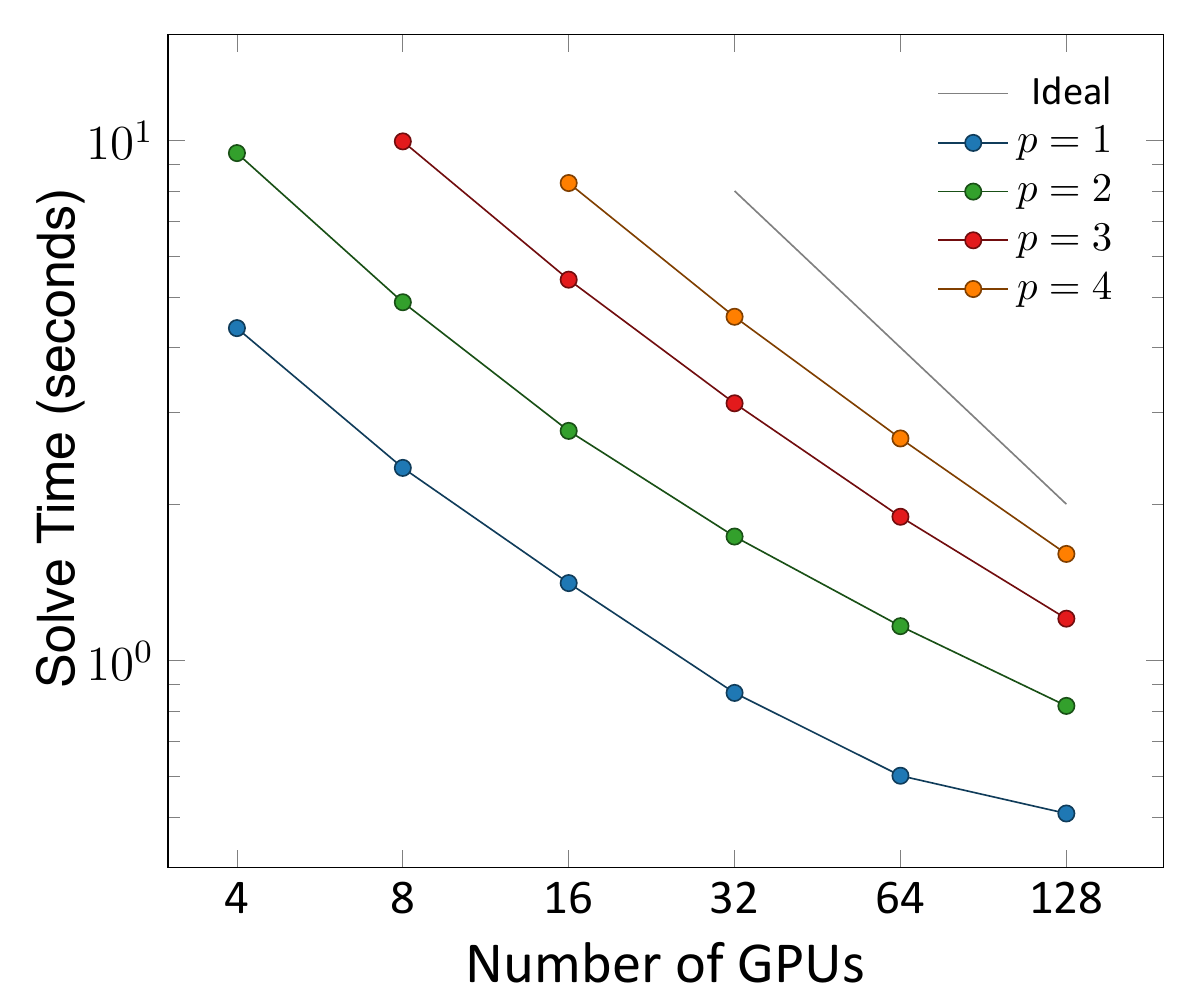}}
  \sisetup{detect-weight=true,mode=text}
  \newcolumntype{Y}{S[table-format=3.1]}
  \setlength{\tabcolsep}{10pt}
  \begin{tabular}{rYYYY}
  \toprule
  & \multicolumn{4}{c}{Time to mesh optimization (sec.)}\\
  & {$p = 1$} & {$p = 2$} & {$p = 3$} & {$p = 4$}  \\
  \midrule
  CPU & 18.8  & 43.0 & 129.6 & 224.3  \\
  GPU & 0.4  & 1.0 & 3.9  & 7.5 \\
  \midrule
  Speedup & \textbf{47$\times$} & \textbf{43$\times$} & \textbf{33$\times$} & \textbf{30$\times$}\\
  \bottomrule
  \end{tabular}
  \caption{Throughput for action of the TMOP Hessian $\partial ^2 F$ (left plot) and strong-scaling of the total time for mesh optimization on GPUs (center plot).
  Table: Comparison of total time for mesh optimization for meshes of different polynomial degrees.}
  \label{fig_tmop_gpu}
\end{figure}

Finally, to support MFEM-based applications that leverage GPUs, we have
implemented TMOP-related operators with sum-factorization and partial
assembly \cite{Camier2023}. This includes new GPU kernels for evaluating the
TMOP objective function \eqref{eq_F_full} along with its first
and second-derivatives, and a Jacobi preconditioner for
accelerating the Newton solve.  Use of sum-factorization and partial
assembly on GPUs has significantly reduced the time for Newton solve in
comparison to the traditional fully assembled matrix-based techniques on CPUs.
\Cref{fig_tmop_gpu} and \Cref{fig_tmop_gpu} show results for
the total mesh optimization time on GPU and CPU from a benchmark problem in
\cite{Camier2023}.

The use of TMOP algorithms is demonstrated in the \textit{Mesh Optimizer} and \textit{Mesh Fitting} miniapps, which use the core TMOP functions for CPUs GPUs.
 
\subsection{Submesh interface}
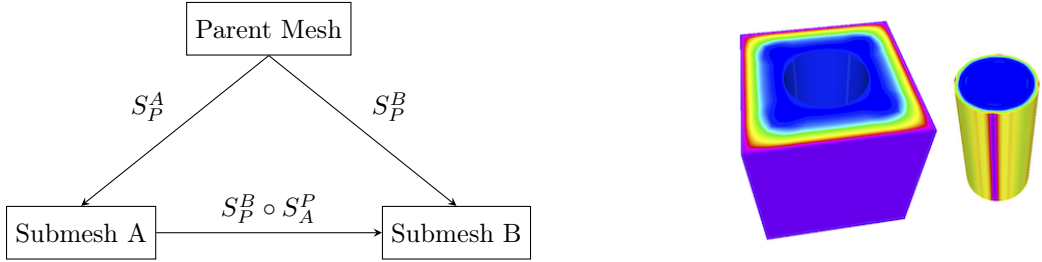
\begin{figure}
    \begin{minipage}{0.5\linewidth}
        \centering
        \begin{tikzpicture}
            [mynode/.style={rectangle, draw=black, minimum size=7mm}]
            \node[mynode] (submesh_a) {Submesh A};
            \node[mynode, right=3cm of submesh_a] (submesh_b) {Submesh B};
            \node[mynode, above=2cm of $(submesh_a.north)!0.5!(submesh_b.north)$] (parent) {Parent Mesh};

            \draw[-stealth] (parent.south) -- (submesh_a.north) node[midway,swap,auto] {$S_P^A$};
            \draw[-stealth] (parent.south) -- (submesh_b.north) node[midway,auto] {$S_P^B$};
            \draw[-stealth] (submesh_a.east) -- (submesh_b.west) node[midway,above] {$S_P^B \circ S_A^P $};
        \end{tikzpicture}
    \end{minipage}\begin{minipage}{0.5\linewidth}
        \centering
        \includegraphics[width=0.5\linewidth]{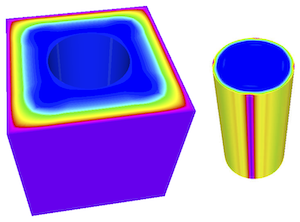}
    \end{minipage}
    \caption{
        Left panel: tree structure showing two submeshes which are derived from the same parent and share a boundary.
        Operations that transfer grid functions between the different submeshes are also illustrated.
        Right panel: Simulation result of first-order coupling between a heat equation and a convection-diffusion equation on submeshes sharing boundary conditions.}
    \label{fig_submesh}
\end{figure}

A recent addition to MFEM is its submesh interface, which allows applications to define a mesh that represents a subset of a parent mesh.
Subsets can be formed from volume elements or boundary elements, creating domain or surface meshes, respectively.
A submesh is a fully functional MFEM mesh, and so all of MFEM's existing machinery can be used on one or multiple submeshes.
Furthermore, MFEM provides the ability to transfer grid functions between a parent mesh and its submeshes or between submeshes that share the same parent Mesh.
Let $S_P^A$ denote the operator that extracts a submesh grid function $u_A$ from the parent grid function $u_p$,  such that $u_A = S_P^A u_p$. We can use this operator to also extend  submesh grid functions to the full mesh and to exchange data between submeshes. See~\Cref{fig_submesh} for an illustration.
This concept can be used to express multiphysics problems in a straightforward way in MFEM.
Another use case for submesh capability is the restriction of operations that compute quantities of interest to specific parts of the domain.

The \textit{Multidomain} miniapp in MFEM is a simple demonstration of how to solve two PDEs, each representing different physics, on the same domain.
MFEM's submesh interface is used to compute on and transfer between the spaces of predefined parts of the domain.
In this miniapp, the spaces on each domain are using the same order $H^1$ finite elements; this is not a requirement, and submesh easily handles different finite element spaces on different domains.
A 3D domain comprised of an outer box with a cylinder shaped inside is used.
A heat equation is described on the outer box domain
\begin{subequations}
\begin{align*}
    \frac{\partial T}{\partial t} & = \kappa \nabla^2 T \, & \mbox{ in outer box},              \\
    T                             & = T_{wall} \,          & \mbox{ on outside wall},           \\
    \nabla T \cdot \vec{n}        & = 0 \,                 & \mbox{ on inside (cylinder) wall},
\end{align*}
\end{subequations}
with temperature $T$, outward unit normal $\vec{n}$ and coefficient $\kappa$.
A convection-diffusion equation is described inside the cylinder domain
\begin{subequations}
\begin{align*}
    \frac{\partial T}{\partial t} & = \kappa \nabla^2 T - \alpha \nabla \cdot (\vec{b} T) \, & \mbox{ in inner cylinder}, \\
    T                             & = T_{wall} \,                                            & \mbox{ on cylinder wall},  \\
    \nabla T \cdot \vec{n}        & = 0 \,                                                   & \mbox{ else},
\end{align*}
\end{subequations}
with temperature $T$, outward unit normal $\vec{n}$, coefficients $\kappa$, $\alpha$ and prescribed velocity profile $\vec{b}$.
To couple the solutions of both equations, a segregated solve with a one way coupling approach is used. The heat equation of the outer box is solved from the time step $T_{box}(t)$ to $T_{box}(t+dt)$.
Then, for the convection-diffusion equation, $T_{wall}$ is set to $T_{box}(t+dt)$ and the equation is solved for $T(t+dt)$, resulting in a first-order one-way coupling.
A visualized result can be seen in~\Cref{fig_submesh}.
 
\section{Applications}
The modular structure of MFEM allows it to powers a wide variety of applications
in areas such as compressible and incompressible flow, electromagnetics,
magnetic and inertial confinement fusion, additive manufacturing, topology
optimization, structural mechanics, subsurface flow, hearth and MRI modeling and
more, see \cite{Anderson2020}. To aid with the development of new applications,
a large number of examples and miniapps are included with the library. In this
section we review some recent MFEM examples, miniapps and application and
describe the PyMFEM interface which allows Python users to take full advantage
of MFEM's capabilities.

\subsection{Random fields and fractional stochastic PDEs}
MFEM's modular and performant implementation enables the design of fast and
scalable solvers for fractional and stochastic PDEs. Such equations arise
naturally in the context of random fields, which are frequently used to model
spatially correlated uncertainties in computational science and engineering
applications.

Whittle \cite{Whittle1954,Whittle1963} first realized that the solution of the
stochastic, fractional PDE
\begin{equation}\label{eq:spde}
 (-\Delta + \kappa) ^ {\alpha/2} u = \mathcal{W}
 ~~
 \text{in~} \Omega
 \,,
\end{equation}
is a Gaussian random field of Mat\'ern covariance. Here, $\mathcal{W}$
denotes spatial Gaussian white noise, while the fractional exponent $\alpha>\operatorname{dim}(\Omega)/2$
and the parameter $\kappa \geq 0$ define the regularity and correlation length, respectively.
Lindgren et al.~\cite{Lindgren2011,Lindgren2022} later popularized using finite elements to solve
\eqref{eq:spde}, calling the approach the \textit{SPDE method}.

The \textit{SPDE} miniapp provides a scalable solver
for \eqref{eq:spde}. It uses a new
implementation of the white noise
sampling proposed in \cite{Croci2018} to handle the stochastic load. It further
employs the triple-A algorithm \cite{Nakatsukasa2018} to construct a rational approximation of the fractional PDE by a set
of integer-order PDEs~\cite{Harizanov2018,Bolin2019}. The
solver avoids repeated matrix assemblies and solves the associated linear
systems with a preconditioned conjugate-gradient algorithm.
The full numerical scheme is described in \cite{duswald2023}.

\begin{figure}[bt!]
   \centering
   \includegraphics[height=3.25cm]{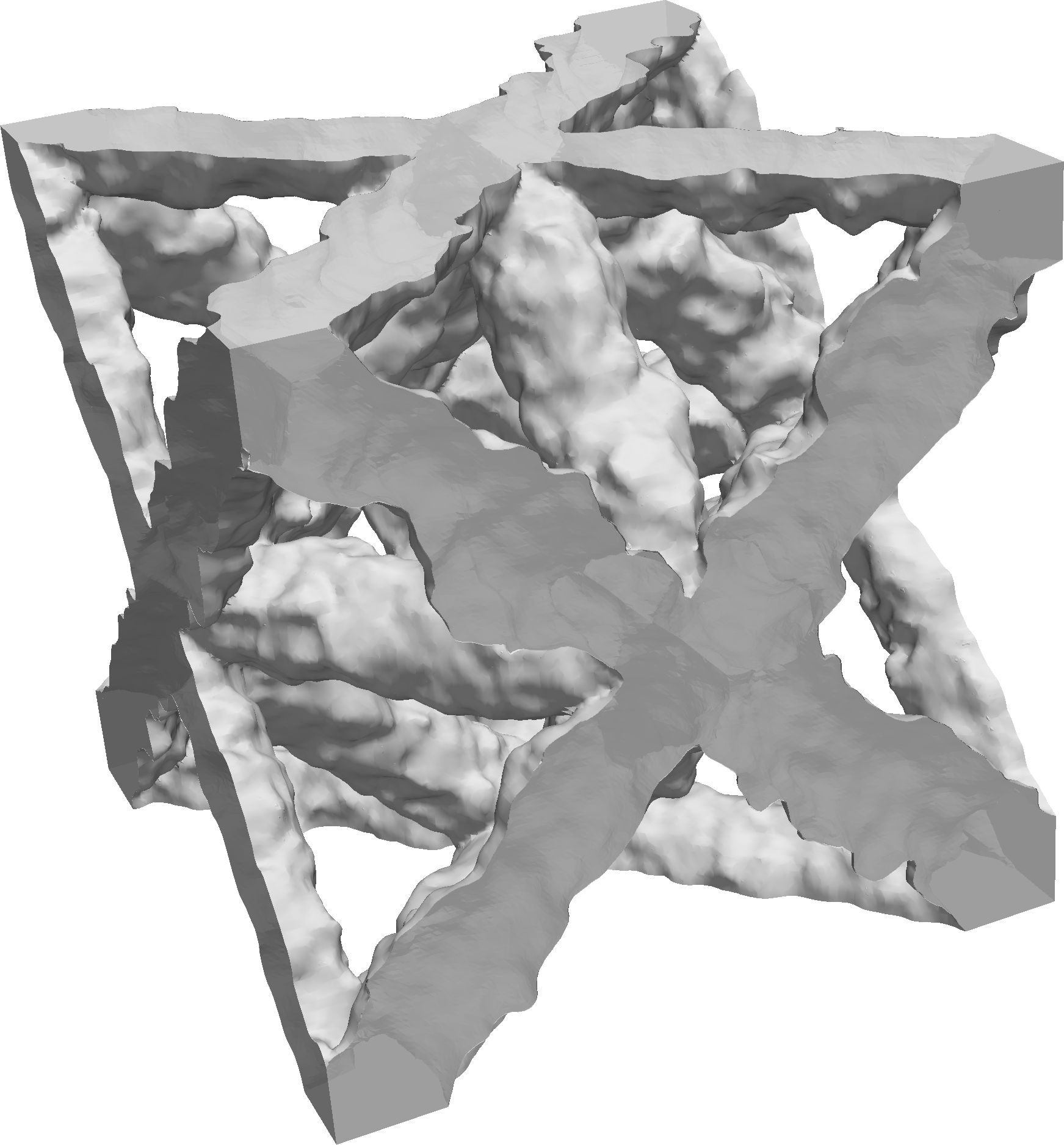}
   \hspace{0.5cm}
   \includegraphics[height=3.25cm]{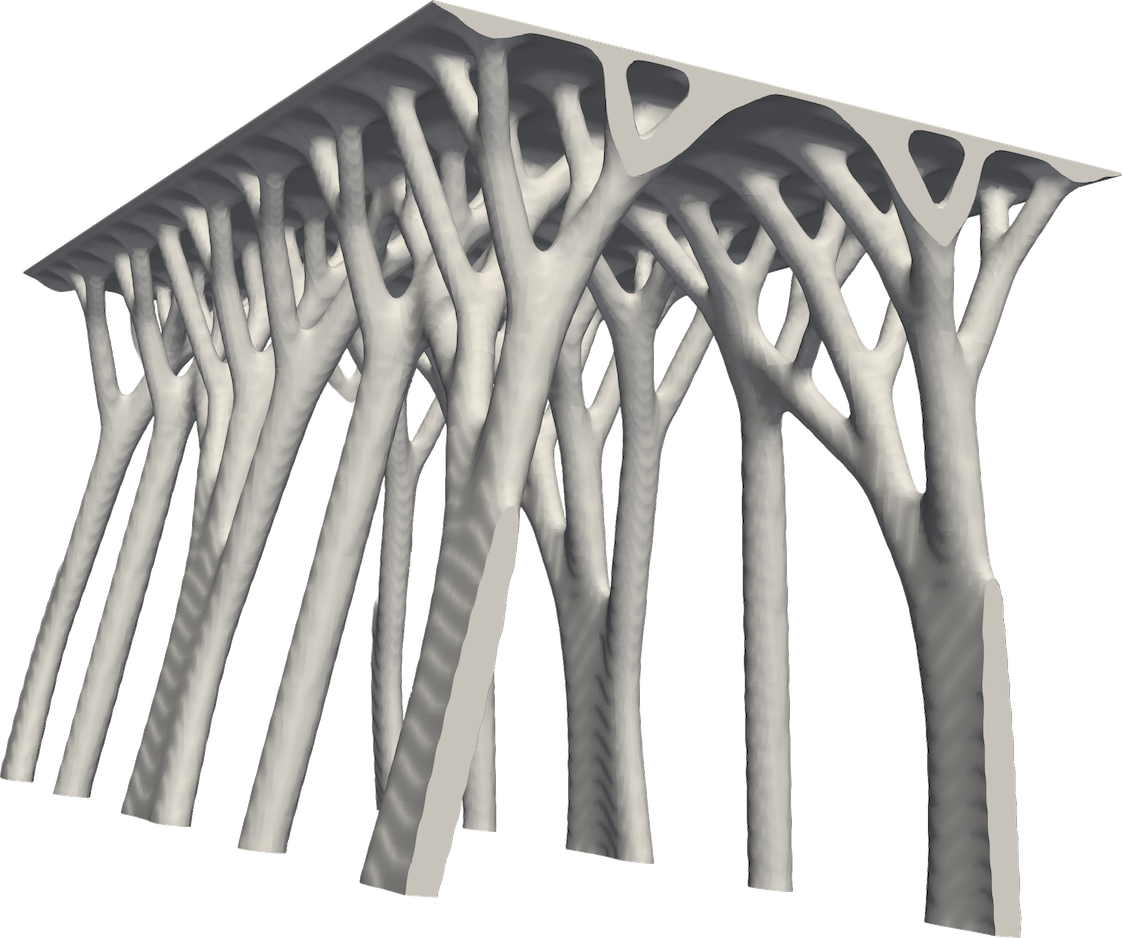}
   \caption{
      Applications of random fields and the SPDE miniapp.
      Left panel: octet-truss: modeling uncertainties in additive manufacturing.
      Right panel: 3D-bridge structure: optimal topology to support an uncertain load~\cite{duswald2023}.
   }
   \label{fig:random-field}
\end{figure}

The SPDE-solver is easy to integrate into MFEM-derived projects and may be used
to model spatial uncertainties with Mat\'ern-type random fields by adding
only two lines of code. The fields subsequently enter into PDE
coefficients, the load, or even distort the domain to describe different types
of uncertainties (e.g., material, geometric, or environmental uncertainties).
Typically, the mathematical model, together with their uncertain SPDE
components, is embedded in workflows such as forward uncertainty quantification
problems or stochastic inverse problems. Examples ranging from biomechanics to
topology optimization under uncertainties are presented in \cite{duswald2023,bollapragada2023adaptive}.
Figure~\ref{fig:random-field} shows a few example applications.

\subsection{Hyperbolic conservation laws}

A new nonlinear integrator for a general system of first-order hyperbolic conservation laws has been introduced. This integrator implements the element-wise weak divergence and the interface flux,
\begin{equation*}
    \sum_{E\in\mathcal{M}}\int_E F(\bm{u}):\nabla \bm{v} \, d\bm{x},~~\sum_{e\in\mathcal{E}}\int_e\widehat{F}(\bm{n}_e;\bm{u}^\pm)\cdot[\![\bm{v}]\!]\, d\bm{x}
    \,,
\end{equation*}
where $F$ is the flux function, and $\widehat{F}$ is the numerical flux function defined on element interfaces.
To implement a specific system, users can create a derived class that provides the action of the flux function.
The user also specifies the choice of numerical flux function $\widehat{F}$.
The Rusanov (local Lax--Friedrichs) flux is provided by MFEM, but other choices of (approximate) Riemann solvers are straightforward to implement by the user.
These new capabilities are illustrated for the Euler equations of gas dynamics in an updated example included with MFEM (example 18).
\Cref{fig_remhos} depicts a snapshot of a solution to the shallow water equations computed using the new integrator.
 
\subsection{High-order ALE}
\begin{figure}
   \centering
   \includegraphics[height=0.15\textwidth]{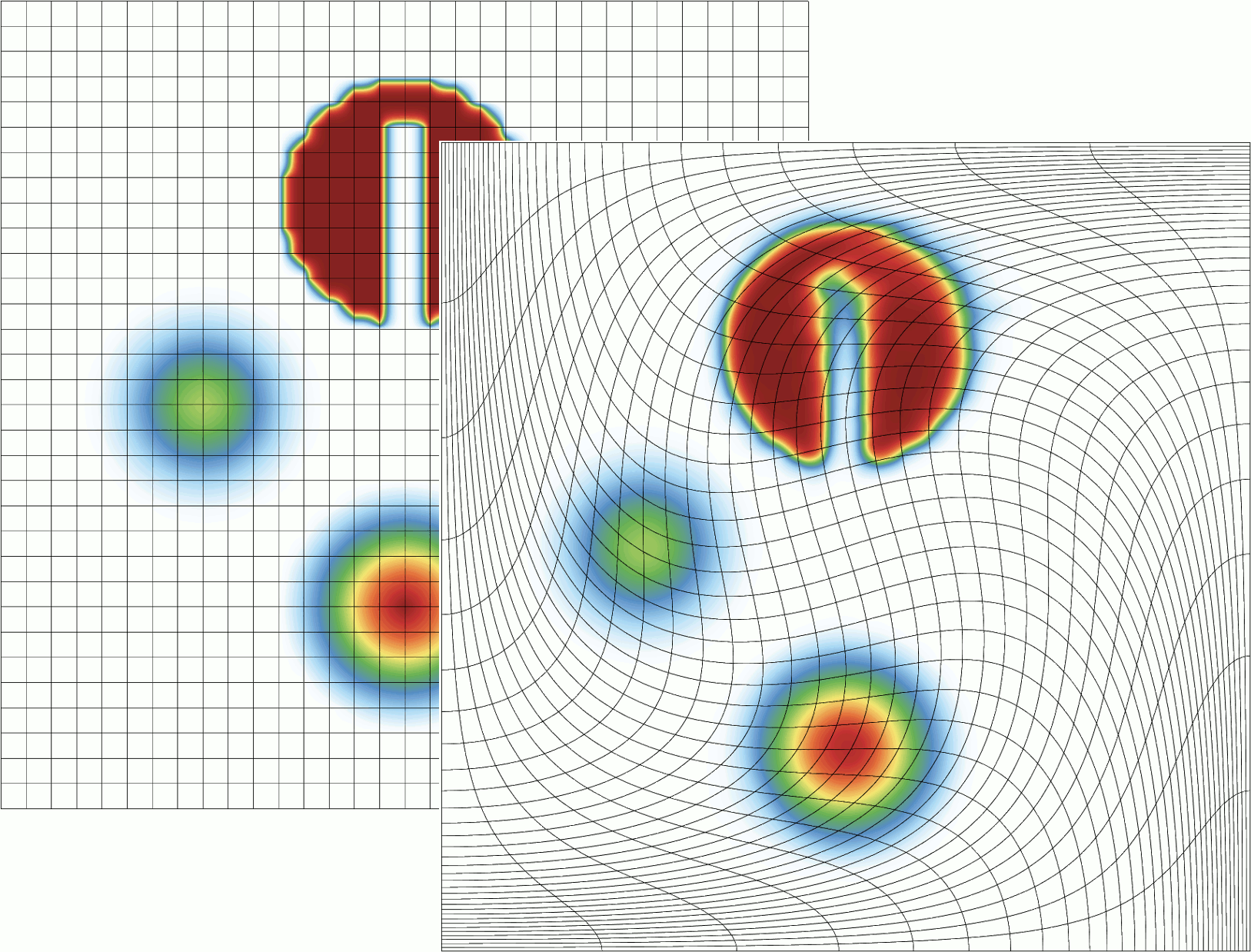}
   \hspace{0.5cm}
   \includegraphics[height=0.17\textwidth]{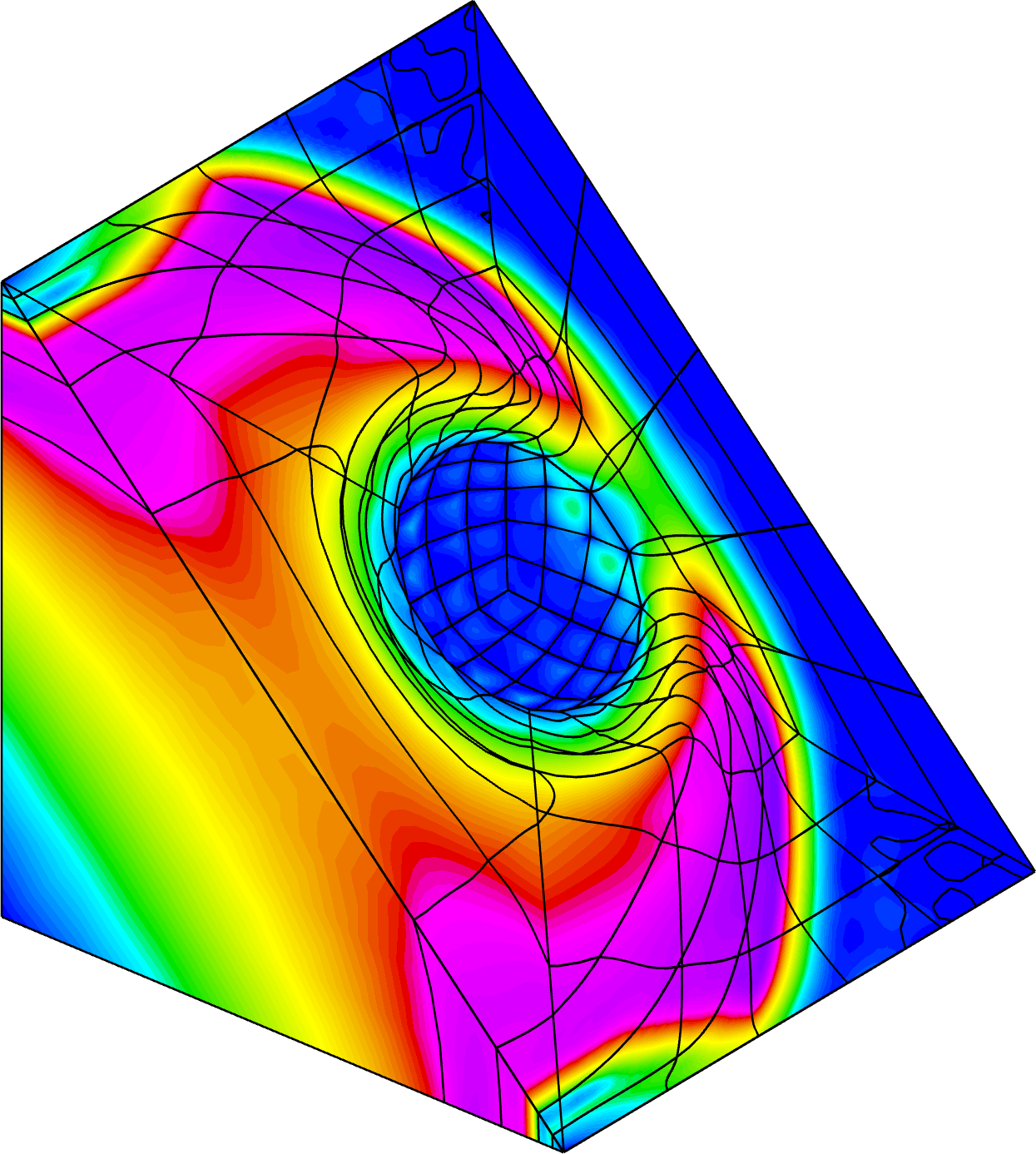}
   \includegraphics[clip=true, trim=0 1cm 0 0.5cm,height=0.17\textwidth]{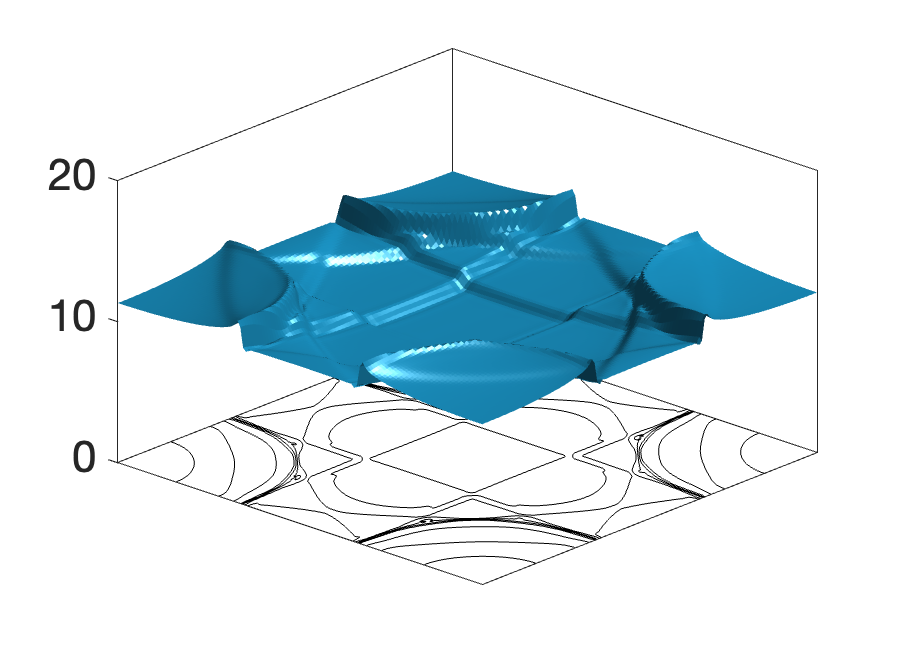}
   \caption{
      Left: remap computation in Remhos.
      Center: simulation of explosion inside a cube with a spherical hole in Laghos.
      Right: The height of a shallow water wave on a periodic square domain.
   }
   \label{fig_remhos}
\end{figure}

Multi-physics Arbitrary Lagrangian-Eulerian (ALE) applications, such as LLNL's MARBL
code \cite{Rieben2020}, are particularly well-suited for using MFEM.
One miniapp developed as part of the MARBL/MFEM collaboration is the
\textit{Remhos} miniapp \cite{Remhos}.
Remhos solves the advection equations that are used to perform monotonic
and conservative discontinuous field interpolation (remap) as part of the
Eulerian phase in ALE simulations
\cite{Dobrev2015, DeLuna2017, Hajduk2020, Kuzmin2020},
see left plot of Figure \ref{fig_remhos}.
The team also continued its research in Lagrangian compressible hydrodynamics
through the Laghos miniapp \cite{Laghos}.
Namely, a novel Nitsche-type approach to weakly enforce free-slip wall
boundary conditions on curved boundaries was developed (see \cite{Atallah2023} and \Cref{fig_remhos}).
Furthermore, a weighted Shifted Interface Method (WSIM) was developed, a new
immersed method for maintaining exact representation of curved interfaces on
high-order Lagrangian grids \cite{Atallah2023immersed}.

\begin{figure}
\begin{center}
\includegraphics[height=0.16\textwidth]{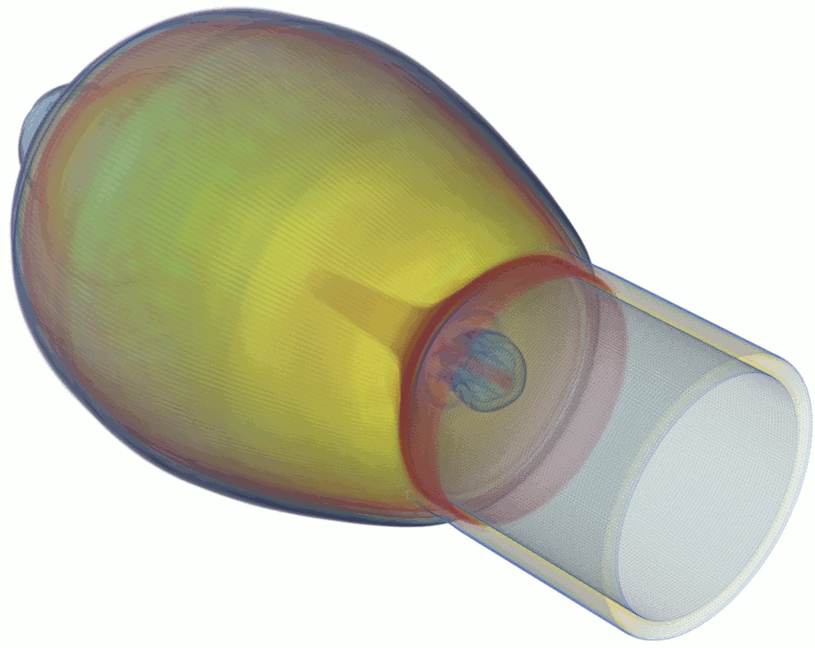} ~~
\includegraphics[height=0.18\textwidth]{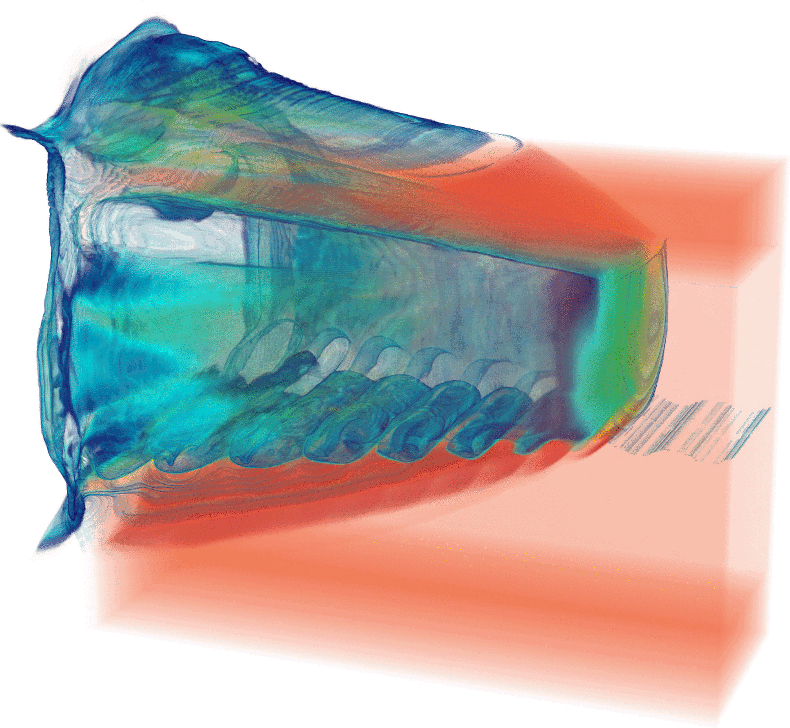}
\end{center}
\caption{Examples of multi-material ALE simulations in MARBL:
         BRL81a shaped charge simulation (left) and
         radiation-driven Kelvin-Helmholtz instability (right).}
\label{fig_marbl}
\end{figure}

LLNL's MARBL code \cite{Rieben2020}, a performance-portable multiphysics
application built on high-order finite elements,
has consistently been on the forefront of leveraging
the latest numerical and algorithmic advancements in MFEM.
Recognizing the performance advantages offered by matrix-free methods, MARBL
has integrated the latest matrix-free algorithms from Laghos, Remhos, and TMOP
to enhance its Lagrange, remap, and mesh optimization phases, respectively.
Moreover, MARBL has leveraged the work of \cite{Pazner2024} to implement
a matrix-free, GPU-accelerated linear solver employing the saddle-point
formulation for solving radiation diffusion equations \cite{Stitt2024}.
These are some of the essential building blocks that have enabled MARBL to
perform fully matrix-free GPU-enabled practical ALE simulations of
multi-material shock and radiation hydrodynamics, like the ones shown in
Figure \ref{fig_marbl}.

\subsection{Electromagnetics applications}
A range of computational electromagnetics applications have been built using MFEM, benefiting from the library's support for adaptive mesh refinement, advanced discretizations, and scalable solvers.

\begin{figure}
\includegraphics[height=0.2\textwidth]{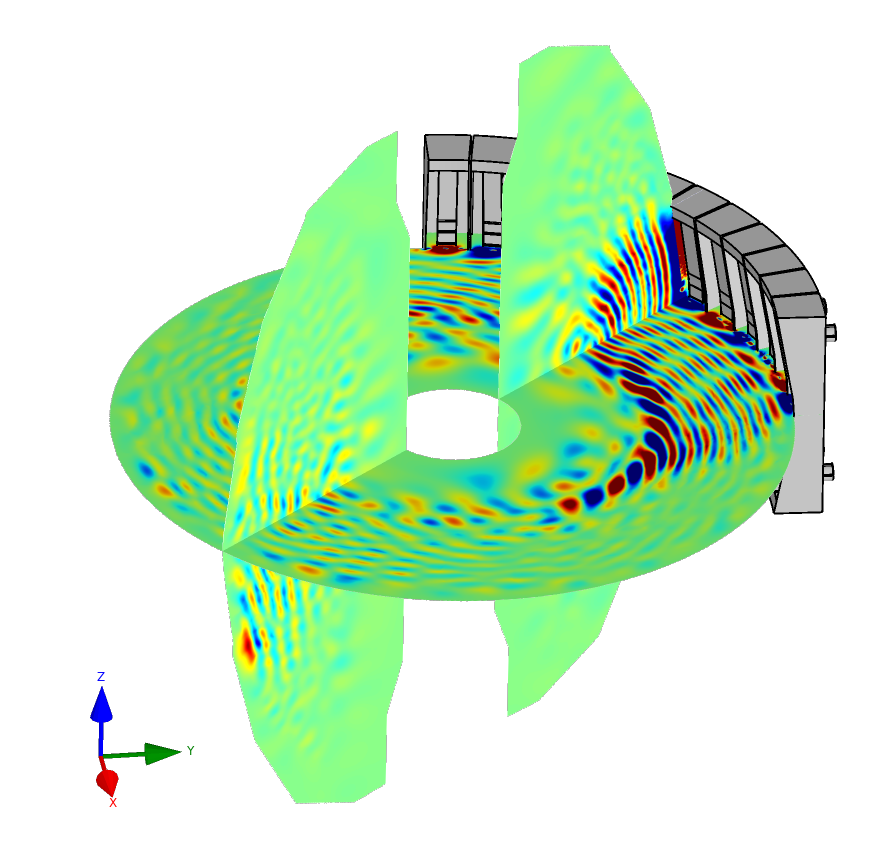}
\includegraphics[height=0.2\textwidth]{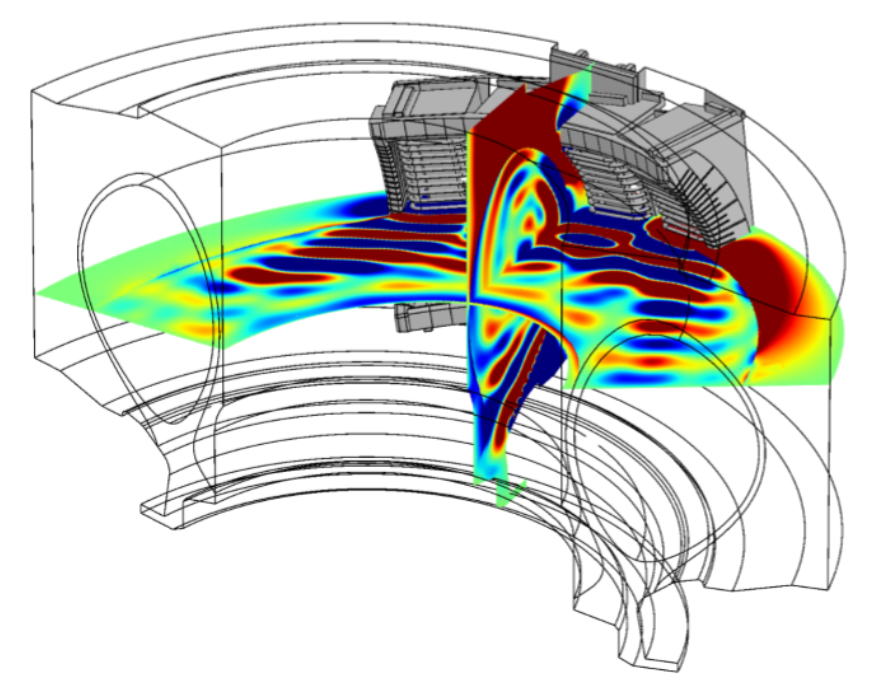}
\caption{Example of RF wave simulations in plasmas. High harmonic fast wave propagation in the NSTX-U spherical tokamak (left), and ion cyclotron radio-frequency waves excited in the Alcator C-Mod tokamak (right).}
\label{fig:hhfw_figs}
\end{figure}

One such application is \textit{Petra-M} (Physics Equation Translator for MFEM).
Petra-M is an open source GUI platform for the finite element method \cite{EPW_Shiraiwa_2017} developed under the Scientific Discovery through Advanced Computing Partnership (SciDAC) program for radio-frequency (RF) wave simulation in fusion plasma (see \Cref{fig:hhfw_figs}).
Petra-M uses MFEM (through the PyMFEM Python interface described in the following section) for its discretization framework and interface to linear solvers.
Its weak form interface allows for the construction of complicated linear system for multiphysics and multidomain problems by selecting MFEM's linear and bilinear form integrators.
Petra-M's GUI interface allows users to build RF simulation using MFEM, and extending Maxwell's system of equations to develop new implementations of physics models, such as radio-frequency sheath \cite{NF_Shiraiwa_2023}, without significant coding efforts.
MFEM's efficient high order discretization allows for the inclusion of the complete 3D magnetic fusion plasmas in RF wave simulations \cite{NF_Bertelli_2022}.

A closely related SciDAC effort is in the process of developing an MFEM-based fluid transport application, called {\textit MAPS} (MFEM Anisotropic Plasma Solver), to simulate the neutrals, ions, and electrons which make up a fusion plasma.
\textit{MAPS} is designed to be coupled with an RF simulator similar to that explored with \textit{Petra-M}.
This coupling will introduce RF heating and pondermotive forces into the fluid equations as well as providing the RF simulator with realistic temperature and density profiles needed to accurately determine the dielectric tensor in Maxwell's equations.
Building upon MFEM as a common code base in these two simulation modules will be crucial for efficiently exchanging accurate field information between these coupled physics models.

MFEM is also the discretization framework for the \textit{Palace} finite element code developed by the Design and Simulation group of the AWS Center for Quantum Computing \cite{Palace}.
Palace is used for performing large-scale, full-wave electromagnetic simulations in both frequency and time domain and makes extensive use of MFEM's solvers, discretizations, and adaptive mesh refinement capabilities.
Palace also uses the libCEED library for performance-portable linear algebra \cite{Brown2021}; libCEED is one of several runtime-configurable backends that MFEM can target for its partial assembly functionality.
 
\subsection{Python interface (PyMFEM)}

MFEM applications can be written using the \textit{PyMFEM} Python interface, in addition to the traditional C++ library API.
Python is a high-level, easy-to-use programming language, widely adopted in educational contexts, and among data scientists and domain scientists.
The development of PyMFEM was initiated to make the MFEM technology more approachable for this broad range of potential users.
In the educational context, PyMFEM allows for the integration of MFEM and its \textit{in situ} visualization tool GLVis into interactive Jupyter notebooks.

PyMFEM facilitates the construction and direct access of MFEM data structure and objects from within Python.
In the current PyMFEM release, all of the included MFEM example programs are translated into Python.
The PyMFEM uses Simple Wrapper Interface Generator (SWIG) for code generation.
The SWIG wrapper recipe files are heavily customized so that a user can interact with MFEM in Python native ways.

PyMFEM also supports using Numba just-in-time compiled Python functions for variable coefficients used in MFEM integrators \cite{Lam2015}.
This is achieved by decorating a Python function with \code{@mfem.jit}.
Numba-based coefficients can have dependencies on other coefficients, including MFEM's native coefficients such as GridFunction coefficients, providing a flexible framework for define complex variable coefficients.
The code snippet in \Cref{fig:pymfem-code} illustrates how such coefficients can be defined.

\makeatletter
\newenvironment{CenteredBox}{\begin{Sbox}}{\end{Sbox}\centerline{\parbox{\wd\@Sbox}{\TheSbox}}}\makeatother

\begin{figure}
    \centering

    \begin{CenteredBox}
        \begin{lstlisting}[language=Python]
@mfem.jit.scalar
def product(ptx):
    return ptx[0] * ptx[1]  # return x*y
# Given a GridFunction p
p_coef = mfem.FunctionCoefficient(p)
@mfem.jit.scalar(depencency=(p_coef,))
def scale(ptx, p_coef):
    return ptx[2]*p_coef  # return z*p
        \end{lstlisting}
    \end{CenteredBox}

    \caption{Illustration of a user-defined coefficient in PyMFEM, using Numba JIT compilation.}
    \label{fig:pymfem-code}
\end{figure}
 
\section{Conclusions}
In this paper we provided a brief summary of some of the recent research, development, and advancements in the open-source MFEM finite element library.
These include state-of-the-art performance on GPU-based supercomputing architectures, high-performance mesh adaptation and optimization, automatic differentiation, support for non-standard discretizations, among many others.
Such features allow MFEM to power a large number of computational physics and engineering applications in areas such as computational electromagnetics, fractional stochastic PDEs, topology optimization, compressible flow, and more.
Further development and improvements to MFEM, including performance optimizations, new discretizations and numerical methods, solvers and preconditioners, miniapps and examples, are continuously underway.
 
\section{Acknowledgements}
This research is supported by the Exascale Computing Project (17-SC-20-SC), a collaborative effort of two U.S. Department of Energy organizations (Office of Science and the National Nuclear Security Administration) responsible for the planning and preparation of a capable exascale ecosystem, including software, applications, hardware, advanced system engineering and early testbed platforms, in support of the nation's exascale computing imperative.
This work was performed under the auspices of the U.S.\ Department of Energy by Lawrence Livermore National Laboratory under Contract DE-AC52-07NA27344.
The work of TD was sponsored by the Wolfgang Gentner Programme of the German Federal Ministry of Education and Research (grant no.\ 13E18CHA).
BK, DK, BL, and SP were partially supported by the LLNL-LDRD Program under Project tracking No.~22-ERD-009.
BK was supported by the U.S.~Department of Energy Office of Science, Early Career Research Program under Award Number DE-SC0024335.
WP was supported by the ORAU Ralph E. Powe Junior Faculty Enhancement Awards Program and NSF RTG DMS-2136228.

\printbibliography

\end{document}